\documentclass[10pt,twocolumn,letterpaper]{article}

\usepackage[pagenumbers]{cvpr}

\usepackage[dvipdfm]{graphicx}
\usepackage{amsmath}
\usepackage{amssymb}
\usepackage{booktabs}

\usepackage[pagebackref,breaklinks,colorlinks]{hyperref}

\usepackage[capitalize]{cleveref}
\crefname{section}{Sec.}{Secs.}
\Crefname{section}{Section}{Sections}
\Crefname{table}{Table}{Tables}
\crefname{table}{Tab.}{Tabs.}

\def\cvprPaperID{2268}
\def\confName{CVPR}
\def\confYear{2022}

\begin{document}

\title{Acquiring a Dynamic Light Field through a Single-Shot Coded Image}

\author{{\fontsize{11pt}{11pt}\selectfont Ryoya Mizuno$^\dagger$, Keita Takahashi$^\dagger$, Michitaka Yoshida$^\ddagger$, 
Chihiro Tsutake$^\dagger$, Toshiaki Fujii$^\dagger$, Hajime Nagahara$^\ddagger$}\\
{\fontsize{11pt}{11pt}\selectfont$^\dagger$Nagoya University, Japan, $^\ddagger$Osaka University, Japan}\\
}


\maketitle

\begin{abstract}
We propose a method for compressively acquiring a dynamic light field (a 5-D volume) through a single-shot coded image (a 2-D measurement). We designed an imaging model that synchronously applies aperture coding and pixel-wise exposure coding within a single exposure time. This coding scheme enables us to effectively embed the original information into a single observed image. The observed image is then fed to a convolutional neural network (CNN) for light-field reconstruction, which is jointly trained with the camera-side coding patterns. We also developed a hardware prototype to capture a real 3-D scene moving over time. We succeeded in acquiring a dynamic light field with 5$\times$5 viewpoints over 4 temporal sub-frames (100 views in total) from a single observed image. Repeating capture and reconstruction processes over time, we can acquire a dynamic light field at $4\times$ the frame rate of the camera. To our knowledge, our method is the first to achieve a finer temporal resolution than the camera itself in compressive light-field acquisition. Our software is available from our project webpage.\footnote{ https://www.fujii.nuee.nagoya-u.ac.jp/Research/CompCam2}
\end{abstract}

\section{Introduction}
\label{sec:intro}
A light field is represented as a set of multi-view images, where dozens of views are aligned on a 2-D grid with tiny viewpoint intervals. This representation contains rich visual information of a target scene and thus can be used for various applications such as 3-D display~\cite{lee2016additive,takahashi2018display}, view synthesis~\cite{zhou2018stereo, mildenhall2019llff}, depth estimation~~\cite{Williem2017,shin18epinet}, synthetic refocusing~\cite{Isaksen2000,ng2005light}, and object recognition~\cite{Maeno2013,Wang2016material}.
The scope of applications will further expand if the target scene is able to move over time. However, a light field varying over time, i.e., a dynamic light field, is challenging to acquire due to the huge data rate, which is proportional to both the number of views and frame rate.

\begin{figure}[t]
\centering
\includegraphics[width=.90\linewidth]{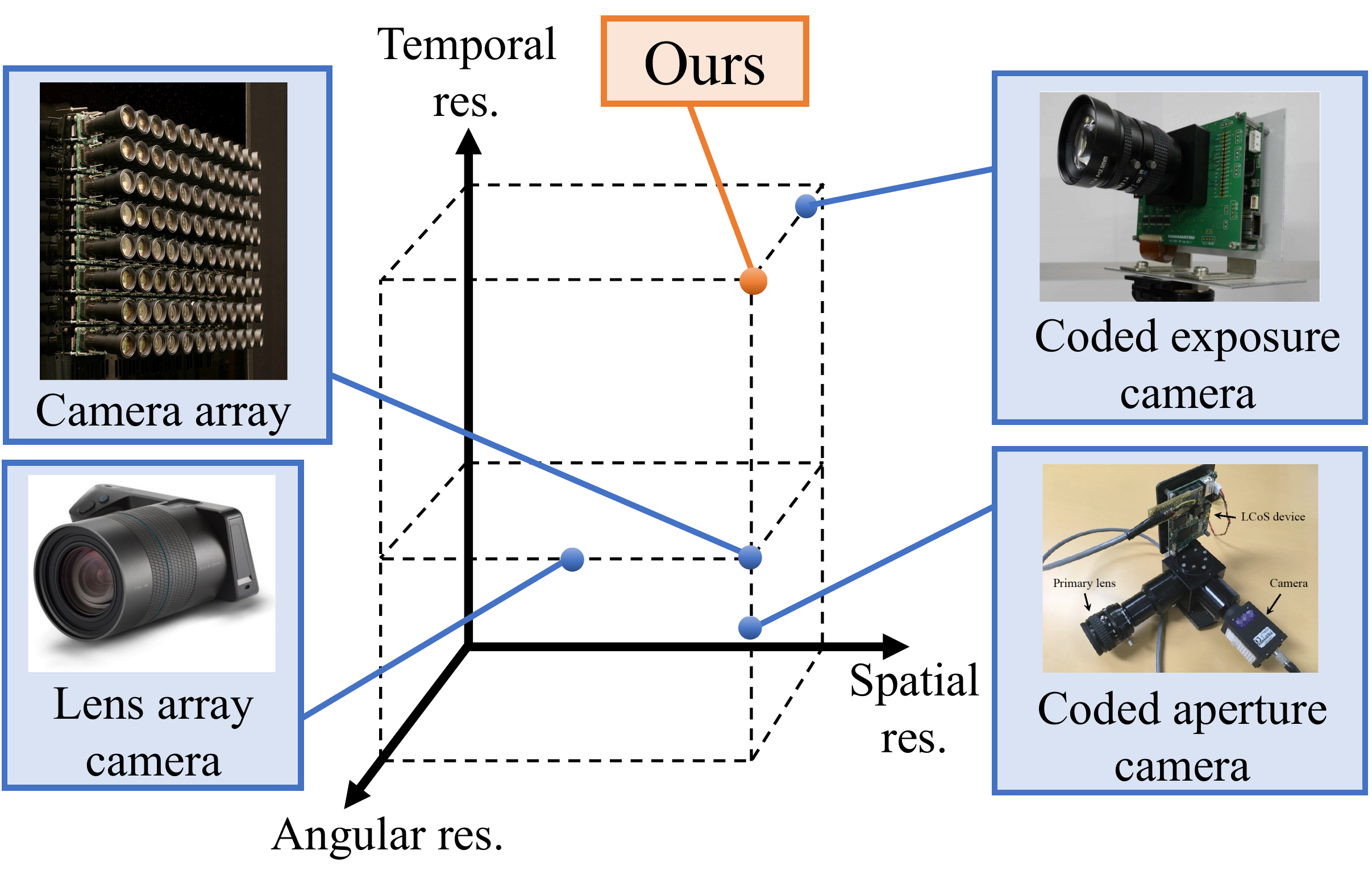}
\caption{Our achievement compared with representative previous works (camera array~\cite{wilburn2005high}, lens-array camera~\cite{ng2006digital}, coded-aperture camera~\cite{Inagaki_2018_ECCV}, and coded exposure camera~\cite{Yoshida_2020}). Axes are in relative scales w.r.t. camera's spatial resolution and frame rate.}
\label{fig:cam_position}
\end{figure}

Several approaches to acquire light fields have been investigated as summarized in Fig.~\ref{fig:cam_position}. The most straightforward approach is to construct an array of cameras~\cite{wilburn2005high,fujii2006multipoint,taguchi2009}, which requires bulky and costly hardware. The second approach is to insert a micro-lens array in front of an image sensor~\cite{adelson1992single,arai1998gradient,ng2005light,ng2006digital,raytrix,Wang2017hybrid}, which enables us to capture a light field in a single-shot image. However, the spatial resolution of each viewpoint image is sacrificed for the angular resolution (number of views). In the above two approaches, the frame rate of the acquired light field is at most equivalent to that of the cameras. Moreover, the data rate is not compressed because each light ray is sampled individually.

The third approach aims to acquire a light field compressively by using a single camera equipped with a coded mask or aperture~\cite{veeraraghavan2007dappled,liang2008programmable,nagahara2010programmable,babacan2012compressive,marwah2013compressive,tambe2013towards,Gupta2017compressive,nabati2018colorLF,Inagaki_2018_ECCV,vadathya2019,Guo_2020_ECCV}. This kind of camera was used to obtain a small number of coded images, from which a light field with the full-sensor spatial resolution can be reconstructed. For static scenes, taking more images with different coding patterns is beneficial to achieve higher reconstruction quality. However, for moving scenes, the use of multiple coded images involves additional complexities related to scene motions. Hajisharif et al.~\cite{Hajisharif_2020_SingleSensor} used a high dimensional light-field dictionary that spanned several temporal frames. However, their dictionary-based light-field reconstruction required a prohibitively long computation time. Sakai et al.~\cite{Sakai_2020_ECCV} handled scene motions by alternating two coding patterns over time and by training their CNN-based algorithm on dynamic scenes. However, the light field was reconstructed only for every two temporal frames (at 0.5$\times$ the frame rate of the camera).

In this paper, we advance the compressive approach several steps further to innovate the imaging method for a dynamic light field. As shown in Fig.~\ref{fig:cam_position}, our method pursues the full-sensor spatial resolution and a faster frame rate than the camera itself. To this end, we design an imaging model that synchronously applies aperture coding \cite{liang2008programmable,nagahara2010programmable,Inagaki_2018_ECCV} and pixel-wise exposure coding~\cite{Reddy_2011_CVPR,hitomi_2011_ICCV,Wei_2018_ECCV,Yoshida_2020} \textit{within} a single exposure time. This coding scheme enables us to effectively embed the original information (a 5-D volume of a dynamic light field) into a single coded image (a 2-D measurement). The coded image is then fed to a CNN for light-field reconstruction, which is jointly trained with the camera-side coding patterns. We also develop a hardware prototype to capture real 3-D scenes moving over time. As a result, we succeeded in acquiring the dynamic light field with 5$\times$5 viewpoints over 4 temporal sub-frames (100 views in total) from a single coded image. Repeating capture and reconstruction processes over time, we acquired a dynamic light field at $4\times$ the frame rate of the camera. To our knowledge, our method is the first to achieve a finer temporal resolution than the camera itself in compressive light-field acquisition.

\section{Background}
\subsection{Computational Photography}
In the literature of computational photography, aperture coding has been used to encode the viewpoint (angular) dimension of a light field \cite{liang2008programmable,nagahara2010programmable,Inagaki_2018_ECCV,Guo_2020_ECCV}, while exposure coding has been adopted to encode fast temporal changes in a monocular video~\cite{Raskar_2006_CodedEP,Reddy_2011_CVPR,hitomi_2011_ICCV,Wei_2018_ECCV,Yoshida_2020}. Our method combines them to encode both the viewpoint (angular) and temporal dimensions simultaneously. Our method is also considered as an extreme case of snapshot compressive imaging~\cite{Wagadarikar_2008,Yuan_2020,Yuan_2021_Snapshot}, where a higher dimensional (typically 3-D) data volume is compressed into a 2-D sensor measurement. 

We noticed that Vargas et al.~\cite{vargas_2021_ICCV} recently proposed an imaging architecture similar to ours for compressive light field acquisition. However, their method was designed for static light fields. Accordingly, their image formation model implicitly assumed that the target light field should be invariant during an exposure time (during the period when the time-varying coding patterns were applied), which is theoretically incompatible with moving scenes. Moreover, they did not report hardware implementation for the pixel-wise exposure coding. In contrast, our method is designed to handle motions during each exposure time, and it is fully implemented as a hardware prototype.

We model the entire imaging pipeline (coded-image acquisition and light-field reconstruction) as a deep neural network, and jointly optimize the camera-side coding patterns and the reconstruction algorithm. This design aligns with the recent trend of deep optics~\cite{Chakrabarti2016learning,Nie_2018_CVPR,sun_2020_ICCP,Iliadis_2016_mask,Li_2020_ICCP,Wu_2019_ICCP,Yoshida_2020,Inagaki_2018_ECCV,Sakai_2020_ECCV} where optical elements and computational algorithms are jointly optimized under the framework of deep learning. However, our method is designed to handle higher dimensional data (dynamic light fields) than the previous works.

\subsection{Light-Field Reconstruction}
Reconstruction of a light field from a coded/compressed measurement is considered as an inverse problem, for which several classes of methods can be used. Traditional methods~\cite{babacan2012compressive,marwah2013compressive,Miandji_2019} formulated this problem as energy minimization with rather simple explicitly-defined prior terms and solved them using iterative algorithms. These methods often result in insufficient reconstruction quality and long computation time. Recently, deep-learning-based methods~\cite{Gupta2017compressive,nabati2018colorLF,vadathya2019,Inagaki_2018_ECCV,Wang_2018_ECCV,Yeung_2019} have gained more popularity due to the excellent representation capability of data-driven implicit priors. Trained on a suitable dataset, these methods can acquire the capability of high-quality reconstruction. Moreover, reconstruction (inference) on a pre-trained network does not require much computation time. Hybrid approaches have also been investigated. Algorithm unrolling methods~\cite{Guo_2020_ECCV,Monga_2021} unroll procedures of iterative algorithms into trainable networks, whereas plug-and-play methods~\cite{Yuan_2020,Yuan_2021_Snapshot} use pre-trained network models as building blocks of iterative algorithms.


We take a deep-learning-based approach and jointly optimize the entire process (coded-image acquisition and light-field reconstruction) in the spirit of deep optics. For the reconstruction part, we use a rather plain network architecture to balance the reconstruction quality and the computational efficiency. Further improvement would be expected with more sophisticated and light-field specific network architectures~\cite{Yeung_2019,Guo_2020_ECCV}. We leave this as future work, because the main focus of this paper is the design of the image acquisition process rather than the reconstruction network.

In recent years, view synthesis from a single image~\cite{Srinivasan2017,single_view_mpi,Niklaus_2019_single,Wiles_2020_CVPR,Shih_2020_3DP20,hu2021worldsheet} has attracted much attention. In principle, 3-D reconstruction/rendering from an ordinary monocular image (without coding) is an ill-posed problem; the results are \textit{hallucinated} by using the implicit scene priors learned from the training dataset rather than the physical cues. In contrast, our method aims to \textit{recover} the 3D and motion information that is \textit{embedded} into a single image through the camera-side coding process.

\section{Proposed Method}
\label{sec:propose}
\subsection{Notations and Problem Formulation}
\label{subsec:problem}
\begin{figure}[t]
\centering
\includegraphics[width=.95\linewidth]{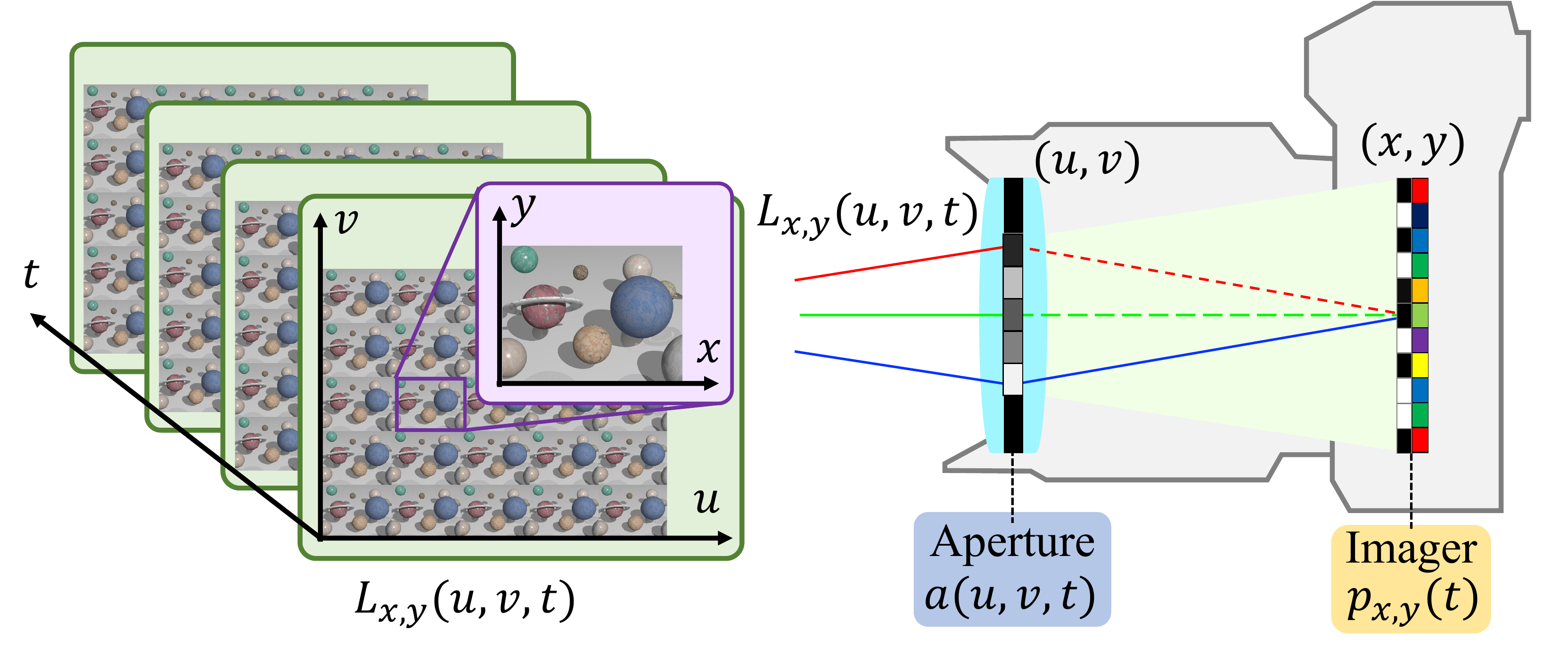}
\caption{Example of dynamic light field (left) and schematic diagram of camera (right).}
\label{fig:camera}
\end{figure}

A schematic diagram of the camera we assume is shown in Fig.~\ref{fig:camera}. Each light ray coming into the camera is parameterized with five variables, $(u,v,x,y,t)$, where $(u,v)$ and $(x,y)$ denote the intersections with the aperture and imaging planes, respectively, and $t$ denotes the time \textit{within} a single exposure time of the camera. We discretize the variable space into a 5-D integer grid, where the range of each variable is described as $S_{\xi} = [0, N_{\xi})$ ($\xi \in \{x,y,u,v,t\}$). By using these variables, the intensity of a light ray is described as $L_{x,y}(u,v,t)$\footnote{For simplicity, we assume that a light field has a single color channel. When handling a light field with RGB colors, we treat each color channel as an individual light field.}. Since $(u,v)$ is associated with the viewpoint (angle), $L_{x,y}(u,v,t)$ is equivalent to a set of multi-view videos, i.e., a dynamic light field.

Our aim is to acquire the latent dynamic light field $L_{x,y}(u,v,t)$:~a 5-D volume with $N_x N_y N_u N_v N_t$ unknowns, from a single coded image $I_{x,y}$:~a 2-D measurement with $N_x N_y$ observables. Hereafter, we assume $N_u = N_v = 5$ and $N_t = 4$ unless mentioned otherwise.

\subsection{Image Acquisition Model}
\label{subsec:principle}
If the camera has no coding functionalities (in the case of an ordinary camera), the observed image is given by
\begin{align}
I_{x,y} = \hspace{-5mm} \sum_{(u,v,t) \in S_u \times S_v \times S_t} \hspace{-7mm} L_{x,y}(u,v,t).
\label{eq:n_cam}
\end{align}
Each pixel value, $I_{x,y}, $ is the sum of light rays over the viewpoint $(u,v)$ and temporal $(t)$ dimensions. Therefore, the variation along $u,v,t$ dimensions is simply blurred out, making it difficult to recover.

Meanwhile, we design an imaging method that can effectively preserve the original 5-D information. We exploit the combination of aperture coding and pixel-wise exposure coding that are synchronously varied \textit{within} a single exposure time. The observed image is given as
\begin{align}
I_{x,y} = \hspace{-5mm} \sum_{(u,v,t) \in S_u \times S_v \times S_t} \hspace{-7mm}  a(u,v,t) \: p_{x,y}(t) \: L_{x,y}(u,v,t).
\label{eq:imaging}
\end{align}
where $a(u,v,t)\in [0,1]$ (semi-transparency) and $p_{x,y}(t)\in\{0,1\}$ (on/off) are coding patterns applied on the aperture and pixel planes, respectively. This imaging process can be regarded as two-step coding as follows. First, a series of aperture coding patterns, $a(u,v,t)$, is applied to $L_{x,y}(u,v,t)$ over time, which reduces the original 5-D volume into a 3-D spatio-temporal tensor, $J_{x,y}(t)$, as
\begin{align}
J_{x,y}(t) = \hspace{-5mm} \sum_{(u,v) \in S_u \times S_v} \hspace{-4mm}  a(u,v,t) \: L_{x,y}(u,v,t). \label{eq:imaging1}
\end{align}
Next, the 3-D tensor, $J_{x,y}(t)$, is further reduced into a 2-D measurement, $I_{x,y}$, through the pixel-wise exposure coding over time using $p_{x,y}(t)$, as
\begin{align}
I_{x,y} = \sum_{t \in S_t} p_{x,y}(t) \: J_{x,y}(t).
\label{eq:imaging2}
\end{align}
By combining these two steps, we encode both the viewpoint ($u,v$) and temporal ($t$) dimensions and embed them into a single 2-D image.

\begin{figure}[t]
\centering
\includegraphics[width=.9\linewidth]{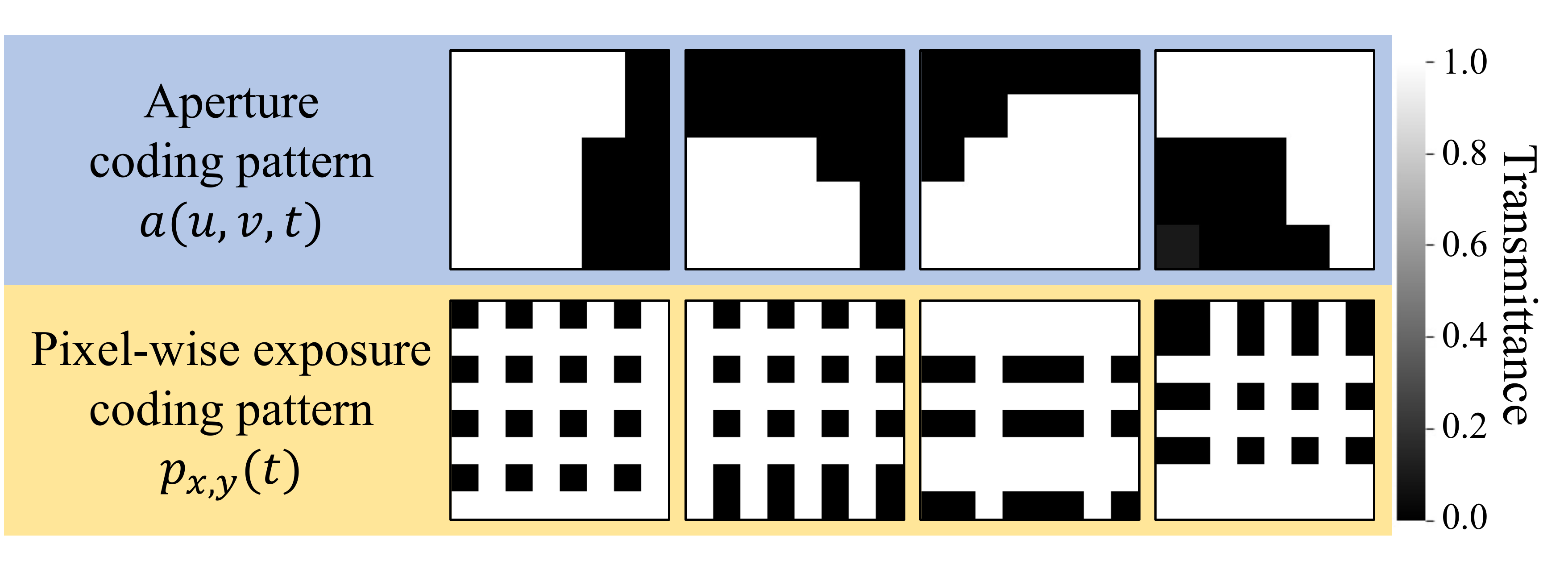}
\caption{Coding patterns applied on aperture and pixel planes.}
\label{fig:mask_pattern}
\end{figure}

An example of the coding patterns is shown in Fig.~\ref{fig:mask_pattern}. As mentioned later, these patterns are directly linked with the parameters of a CNN (AcqNet), which is jointly trained with another CNN for light-field reconstruction (RecNet). Therefore, these coding patterns are optimized for the training dataset so as to preserve as much of the light-field information as possible in the observed image.

\begin{figure}[t]
\centering
\includegraphics[width=.8\linewidth]{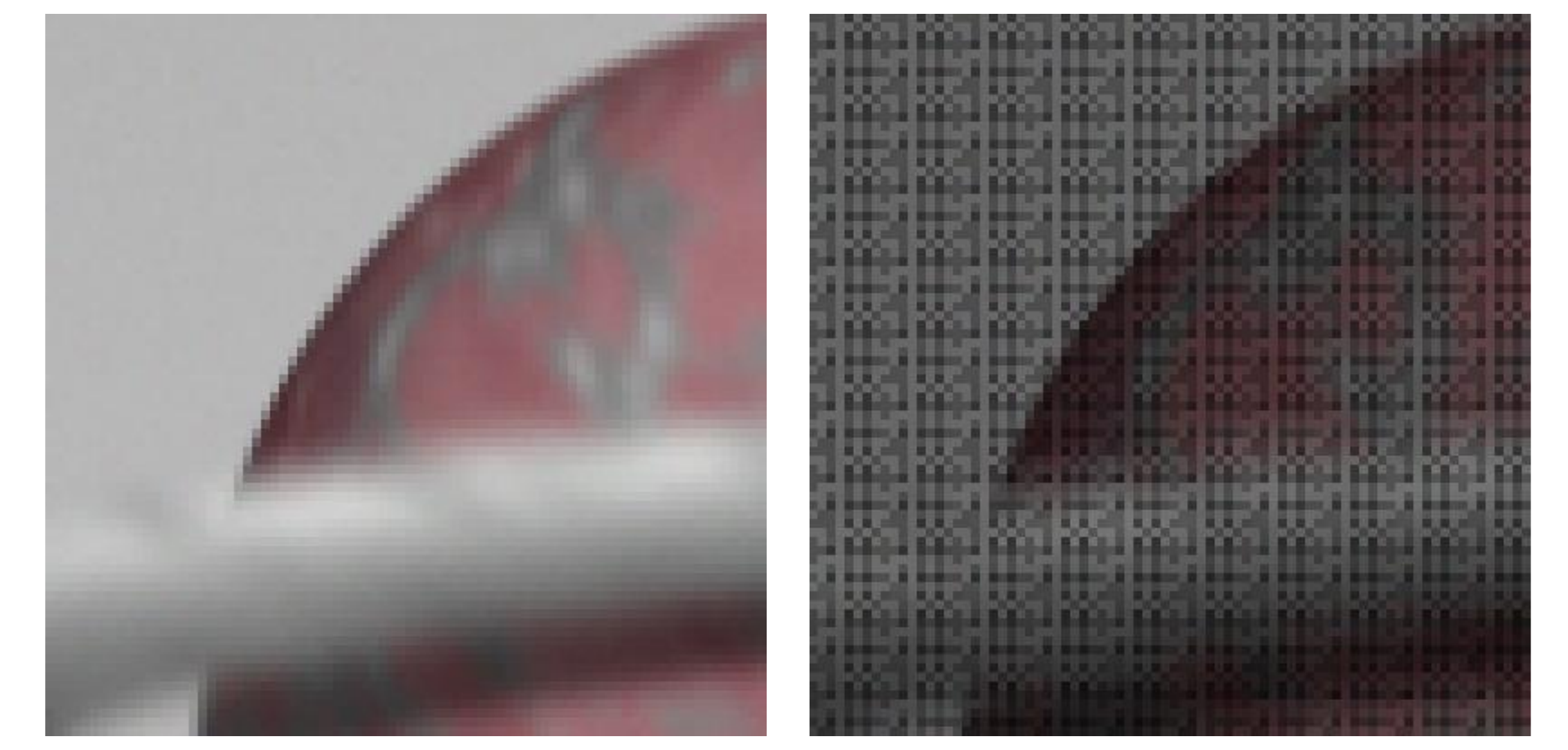}
\caption{Example images acquired by ordinary camera Eq.~(\ref{eq:n_cam}) (left) and our imaging model of Eq.~(\ref{eq:imaging}) (right).}
\label{fig:aq}
\includegraphics[width=.8\linewidth]{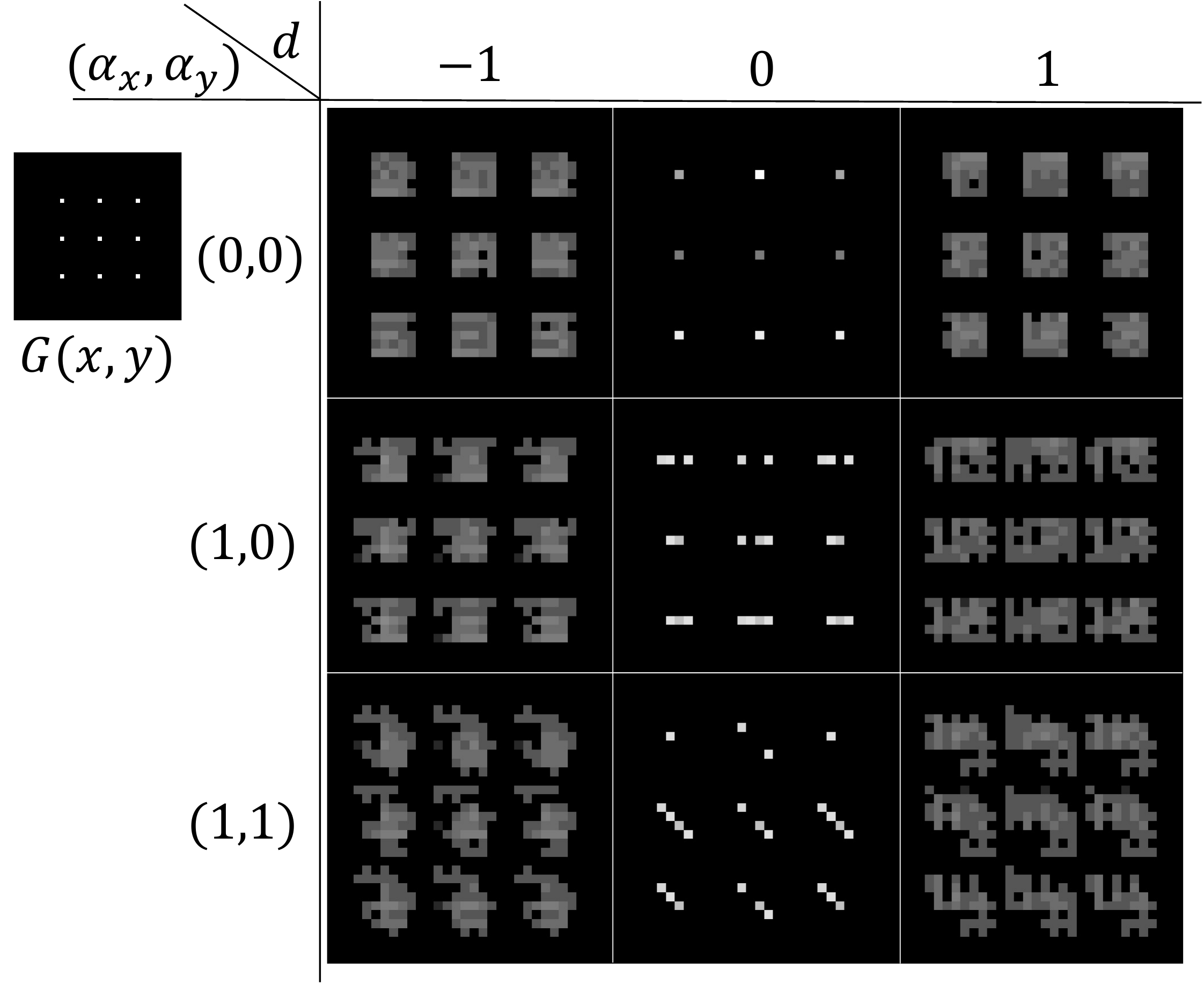}
\caption{Our imaging model yields distinct PSFs for different motion and disparity values (coding patterns in Fig.~\ref{fig:mask_pattern} were used).}
\label{fig:psf}
\end{figure}

Figure~\ref{fig:aq} shows two images (close-ups of the same portion) obtained from a test scene through two imaging models: the ordinary camera (Eq.~(\ref{eq:n_cam})) and ours (Eq.~(\ref{eq:imaging})). The ordinary camera obtains a simply blurred observation, while ours obtains a dappled image due to the coding patterns. To further analyze the effect of coding, we also used a primitive scene with a fronto-parallel plane (a primitive plane scene). As shown in Fig.~\ref{fig:psf}, we prepared an image $G(x,y)$ with nine bright points as the texture for the plane. We then synthesized a dynamic light field using the parameters for the 2-D lateral velocity $(\alpha_x, \alpha_y)$ [pixels per unit time] and disparity $d$ [pixels per viewpoint] (corresponding to the depth) as
\begin{align}
L_{x,y}(u,v,t) = G(x-du-\alpha_xt, y-dv-\alpha_yt)
\label{eq:mklf}
\end{align}
from which we computed an observed image by using Eq.~(\ref{eq:imaging}). Some resulting images obtained with different parameters are shown in Fig.~\ref{fig:psf} (the brightness is corrected for visualization). These images can be interpreted as point spreading functions (PSFs) for various motion and disparity values. Notably, these PSFs are distinct from each other. Moreover, even in a single image, the PSFs for the nine points differ from each other. These results show that both motions and disparities, which are associated with changes along the temporal ($t$) and viewpoint ($u,v$) dimensions, respectively, are \textit{encoded} by the various shapes of PSFs depending on the spatial coordinate $(x,y)$. The encoded information is not human readable, but can be deciphered by the RecNet that is jointly trained with the coding patterns.

\subsection{Hardware Implementation}
\label{subsec:hard_imp}
\begin{figure}[t]
\centering
\includegraphics[width=.38\linewidth]{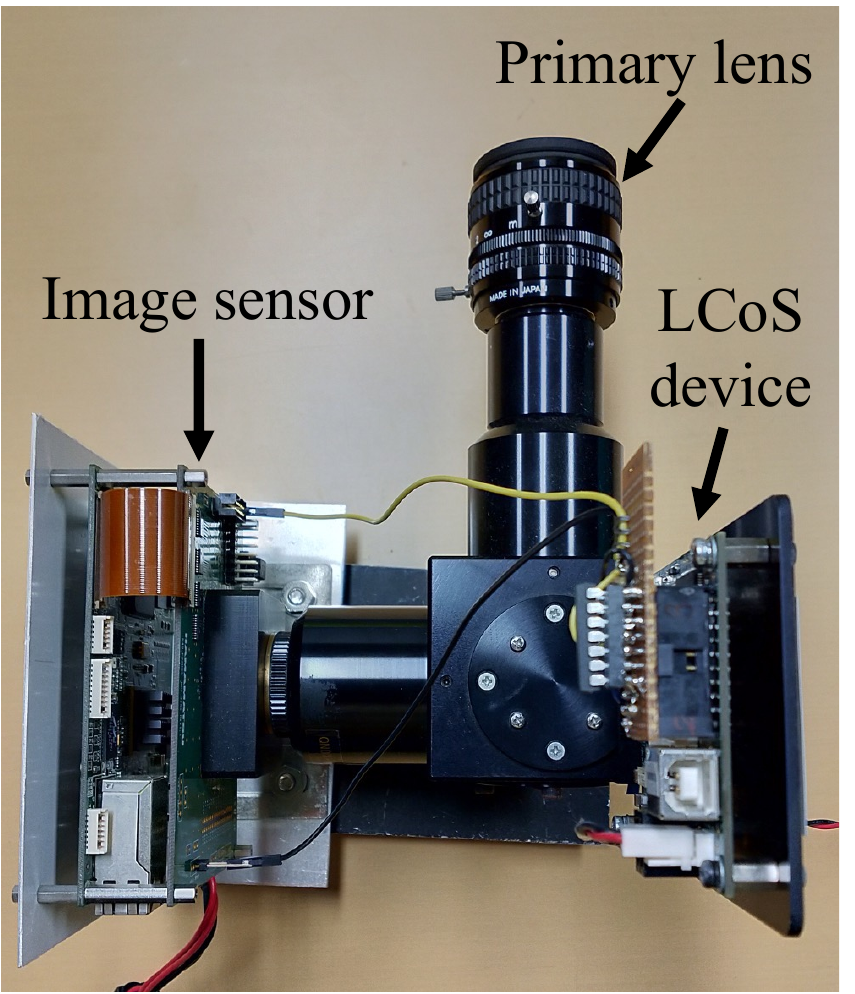}
\includegraphics[width=.45\linewidth]{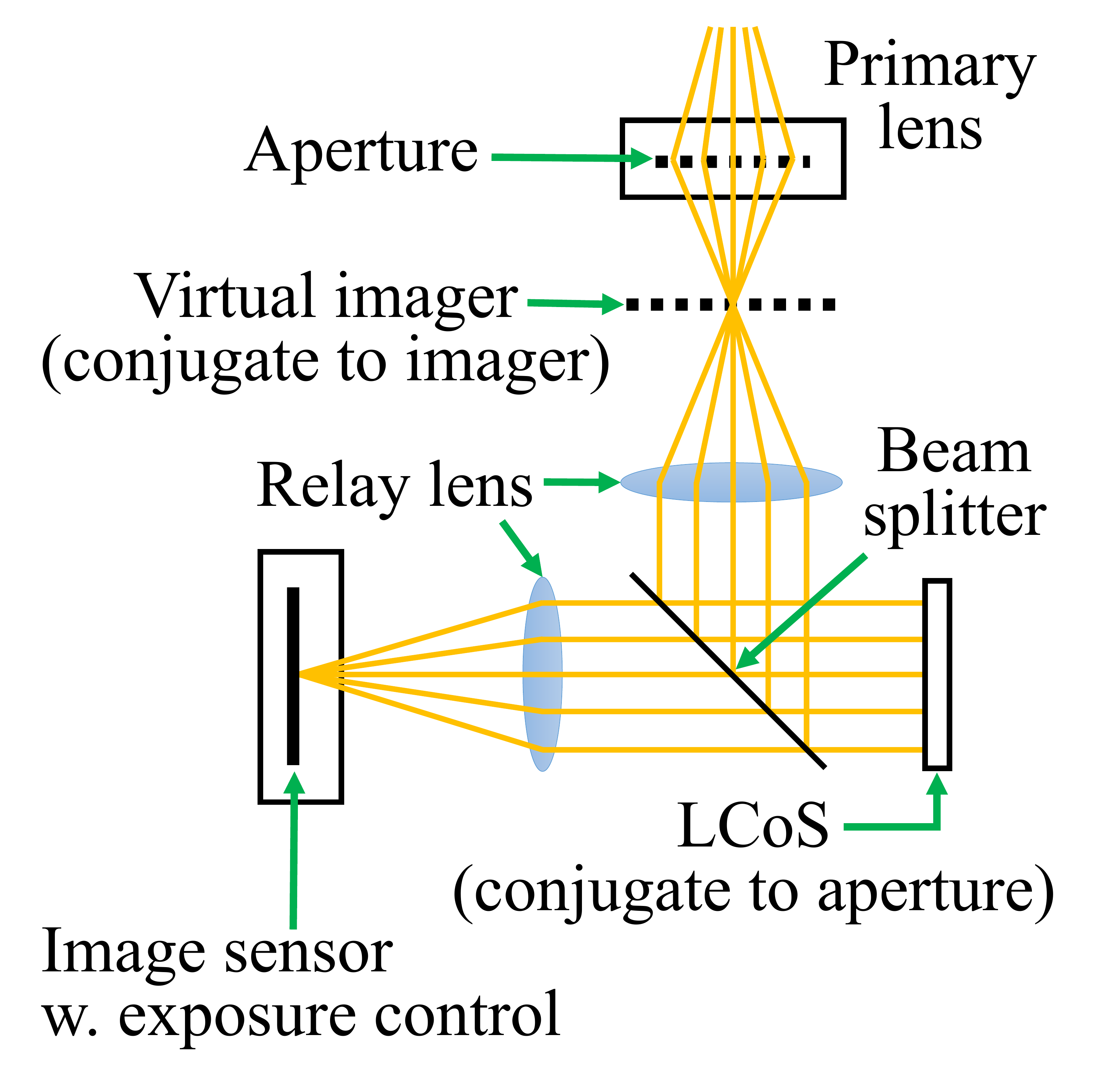}
\caption{Our camera prototype (left) and optical diagram (right).}
\label{fig:camera_optics}
\vspace{2mm}
\includegraphics[width=.85\linewidth]{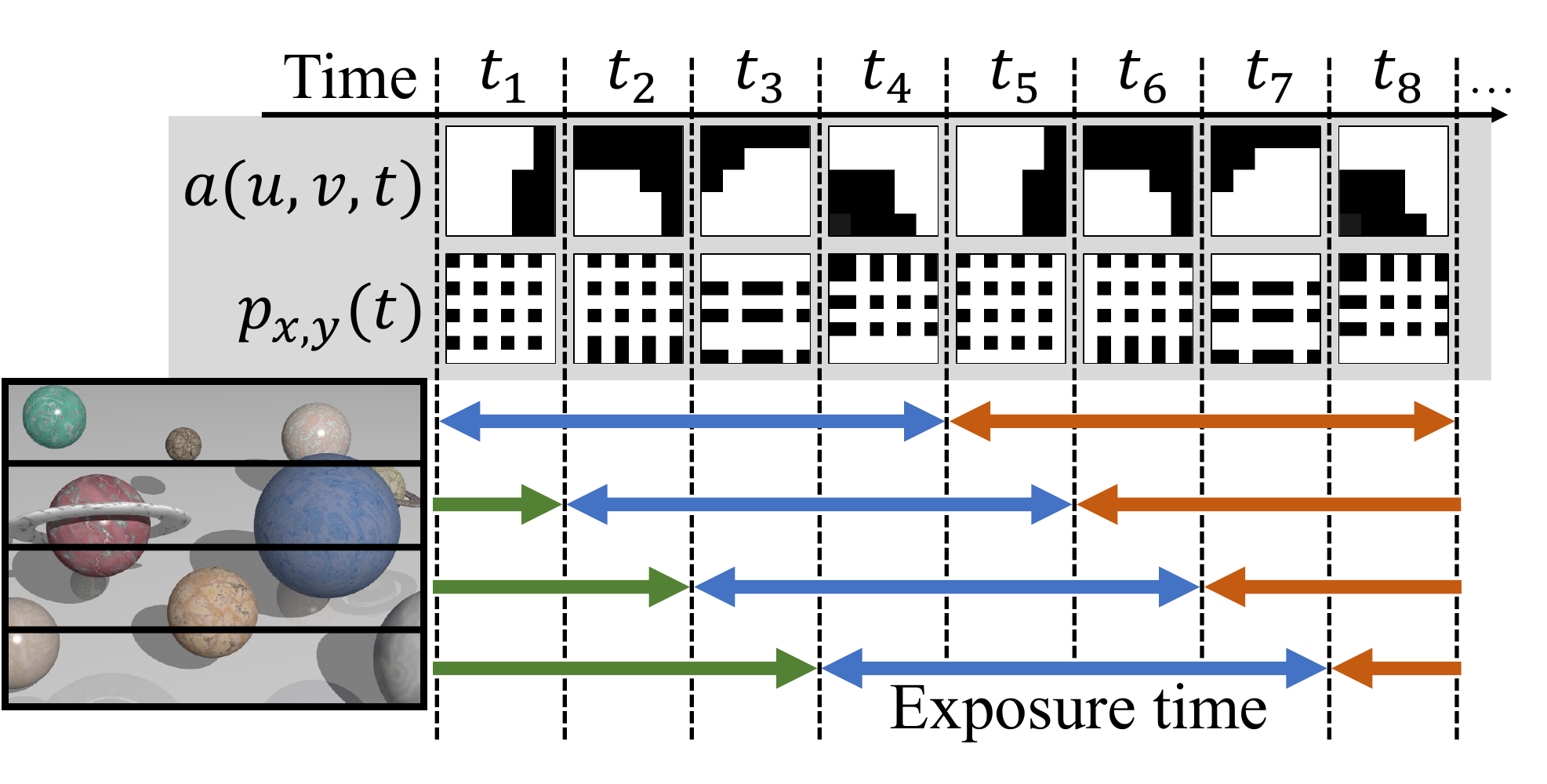}
\caption{Time chart of our camera. Exposure timing is different for four vertically divided regions on image sensor.}
\label{fig:ex_time}
\end{figure}

We developed a prototype camera shown in Fig.~\ref{fig:camera_optics} that can apply aperture coding and pixel-wise exposure coding within a single exposure time. 

We used a Nikon Rayfact (25 mm F1.4 SF2514MC) as the primary lens. The aperture coding was implemented using a liquid crystal on silicon (LCoS) display (Forth Dimension Displays, SXGA-3DM), which had 1280 $\times$ 1024 pixels. We divided the central area of the LCoS display into 5 $\times$ 5 regions, each with 150 $\times$ 150 pixels. Accordingly, the angular resolution of the light field was set to 5 × 5. The pixel-wise exposure coding was implemented using a row-column-wise exposure sensor~\cite{Yoshida_2020} that had 656 $\times$ 512 pixels. We synchronized the LCoS display with the image sensor via an external circuit, so that four sets of coding patterns were synchronously applied within a single exposure time. The timing chart is shown in Fig.~\ref{fig:ex_time}. The time duration assigned for each coding pattern was set to 17 ms. Accordingly, the unit time for the target light field was also 17 ms (58.8 fps). Meanwhile, a single exposure time of the camera ranged over the 4 time units (temporal sub-frames), and thus, the interval between the two exposed images was 68 ms (14.7 fps in terms of the camera's frame rate).

We mention several restrictions resulting from the image sensor's hardware. First, the sensor was not equipped with RGB filters and was thus incapable of obtaining color information. Second, the coding patterns were not freely designable, because they were generated by the column-wise and row-wise control signals repeating for every 8$\times$8 pixels. Therefore, the applicable coding patterns were limited to binary, 8$\times$8-pixels periodic, and row-column separable ones. This restriction was considered in our network design as mentioned later. Finally, due to the timing of the vertical scan, the time duration covered by a single exposed image depended on the vertical position. More precisely, as shown in Fig.~\ref{fig:ex_time}, the image sensor was vertically divided into 4 regions, each of which had a distinctive exposure timing with 17 ms differences from the neighbors. Accordingly, these regions were modulated by the same four sets of coding patterns but in different orders. To accommodate these differences, we used a single instance for AcqNet, but permuted the order of time units in the input light field for the 4 regions, respectively. We prepared 4 instances of RecNet corresponding to the 4 regions and jointly trained them with the coding patterns. This extension required four region-wise reconstruction processes conducted in parallel, but still maintained $\times 4$ finer temporal resolution than the camera.

\begin{figure*}[t]
\centering
\includegraphics[width=.9\linewidth]{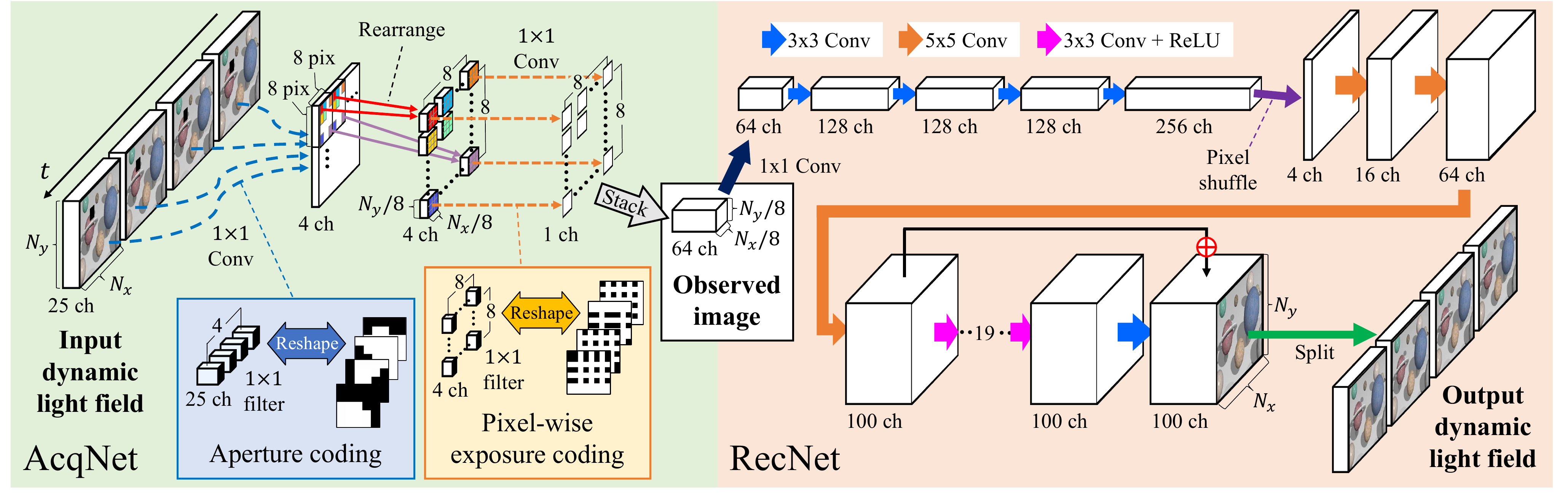}
\caption{Our network architecture consists of AcqNet and RecNet, which correspond to coded image acquisition and light-field reconstruction processes, respectively. Dynamic light-field ranging over four temporal units is processed at once.}
\label{fig:net}
\end{figure*}

\subsection{Network Design and Training}
\label{subsec:net_arc}
As shown in Fig.~\ref{fig:net}, our method was implemented as a fully convolutional network, consisting of AcqNet and RecNet. AcqNet is a differentiable representation of the image formation model with trainable coding patterns, where a target light field is compressed into a single observed image. RecNet was designed to receive the observed image as input and reconstruct the original light field. The entire network was trained end-to-end using the squared error against the ground-truth light field as the loss function. By doing so, the image acquisition and light-field reconstruction processes were jointly optimized. When a real camera was used, the coding patterns for the camera were tuned in accordance with the trained parameters of AcqNet. Then, image acquisition was conducted physically on the imaging hardware, and only the reconstruction (inference on RecNet) was performed on the computer.

AcqNet takes as input a dynamic light field over 4 consecutive time units, which has $N_x \times N_y$ pixels and $5\times 5$ viewpoints over 4 time units. The viewpoint dimensions are unfolded into a single channel, resulting in 4 input tensors with the shape of $25 \times N_x \times N_y$. The first block of AcqNet corresponds to the aperture coding (Eq.~(\ref{eq:imaging1})). To implement this process, we followed Inagaki et al.~\cite{Inagaki_2018_ECCV}; we used 2-D convolutional layers with $1\times1$ kernels and no biases, where each kernel weight corresponds to the apertures' transmittance for each viewpoint. We prepared 4 separate convolutional layers for the 4 time units, in each of which 25 channels were reduced into a single channel. The outputs from these layers are stacked along the channel dimension, resulting in a tensor of $4 \times N_x \times N_y$. The second block corresponds to the pixel-wise exposure coding (Eq.~(\ref{eq:imaging2})), where $8 \times 8$ repetitive patterns are applied. For this process, we prepared 64 separate convolutional layers ($1\times1$ kernels without biases), each of which takes a tensor of $4 \times N_x/8 \times N_y/8$ as input (every $8 \times 8$ pixels extracted from the tensor of $4 \times N_x \times N_y$) and reduces 4 channels into a single channel. To constrain the coding patterns to be hardware implementable (binary and row-column separable), we used the same training technique as Yoshida et al.~\cite{Yoshida_2018_ECCV} (see section 4.1 in~\cite{Yoshida_2018_ECCV}). The outputs from these layers are stacked along the channel dimension, resulting in a tensor of $64 \times N_x/8 \times N_y/8$, which is equivalent to a single observed image with $N_x \times N_y$ pixels. Finally, to account for noise during the acquisition process, Gaussian noise (zero-mean and $\sigma = 0.005$ w.r.t. the range of pixel values $[0,1]$) is added to the observed image.

RecNet accepts an output from AcqNet (or an image acquired from a real camera) as a tensor of $64 \times N_x/8 \times N_y/8$. The first 5 convolutional layers gradually increase the number of channels to 256, while keeping the spatial size unchanged. Then, the tensor is reshaped into $4 \times N_x \times N_y$ using a pixel shuffling operation~\cite{PS}. The subsequent two convolutional layers increase the number of channels to 100, followed by 19 convolutional layers and a residual connection for refinement. The output from RecNet is the latent dynamic light field represented as a tensor of $100 \times N_x \times N_y$, where 100 channels correspond to $5 \times 5$ views over 4 time units (temporal sub-frames). As mentioned in \ref{subsec:hard_imp}, four instances of RecNet should be used in parallel to handle the time differences among the four vertical regions.

We finally mention the training dataset. We first collected 223,020 light-field patches from 51 static light fields with intensity augmentation. Next, following Sakai et al.~\cite{Sakai_2020_ECCV}, we gave 2-D lateral motions (in-plane translations) to the collected patches to synthesize \textit{virtually-moving} light-field samples. We used linear motions with constant velocities: $(\alpha_x, \alpha_y)$ [pixels per unit time], where $\alpha_x, \alpha_y \in \{-2, 1, 0, 1, 2\}$; this is equivalent to at most $\pm$8 pixel translation per frame in terms of the camera's frame rate. This motion model was simple and limited, but it would be sufficient for the motions \textit{within} a single exposure time, which is short enough. We had 25 motion patterns in total, all of which were applied to each light-field patch. To sum up, we had 5,575,500 samples of dynamic light fields, each with $64 \times 64$ pixels at $5 \times 5$ viewpoints over $4$ time units. Note that even a single training sample had a significant size (409,600 elements), which necessitated the network to be lightweight.

We implemented our software using PyTorch. The network was trained over five epochs using the Adam optimizer. The training took approximately seven days on a PC equipped with NVIDIA Geforce RTX 3090. We also trained our model with $8 \times 8$ views and different ranges for the assumed motions $(\alpha_x, \alpha_y)$. Please refer to the supplementary material for details.

\section{Experiments}
\label{sec:experiments}
We conducted several quantitative evaluations using a computer generated scene and experiments using our prototype camera. To summarize, we succeeded in acquiring a dynamic light field with 4$\times$ finer temporal resolution than the camera itself. Note that there is no baseline to compete against, because to our knowledge, no prior works have ever achieved the same goal as ours. Please refer to the supplementary video for better visualization of our results.

\subsection{Quantitative Evaluation}
\label{subsec:quantitative}
\textbf{Ablation study for the coding method}. To validate our image acquisition model in Eq.~(\ref{eq:imaging}), we need to analyze the effect of coding on the aperture ($a(u,v,t)$) and pixel ($p_{x,y}(t)$) planes. In addition to our original method (denoted as \textbf{A+P}), we trained three variants of our methods as follows. \textbf{Ordinary}: no coding was applied ($a(u,v,t) = \mbox{const}$, $p_{x,y}(t) = \mbox{const}$), which corresponded to light-field reconstruction from a single uncoded image. \textbf{A-only}: only the aperture coding was enabled ($p_{x,y}(t)=\mbox{const}$). \textbf{P-only}: only the pixel-wise exposure coding was enabled ($a(u,v,t)=\mbox{const}$). Furthermore, to evaluate the theoretical upper-bound, we also prepared a free-form coding over the 5-D space (denoted as \textbf{Free5D}), given by:
\begin{align}
I_{x,y} = \hspace{-5mm} \sum_{(u,v,t) \in S_u \times S_v \times S_t} \hspace{-7mm}  m(x,y,u,v,t) L_{x,y}(u,v,t)
\end{align}
where $m(x,y,u,v,t) \in [0,1]$ was a fully trainable modulating pattern periodic over $8 \times 8$ pixels. Note that this is only a software simulation; no hardware realization is available. The five methods mentioned so far were different in the imaging models but aimed for the same goal: reconstructing a dynamic light field ($5 \times 5$ views over 4 time units) from a single observed image. For all the methods, RecNets with the same network structure were jointly trained with the respective coding patterns on the same training dataset for the same number of epochs.

\begin{figure*}[t]
\centering
\begin{minipage}[b]{0.16\hsize}
\centering
\includegraphics[width=.95\linewidth]{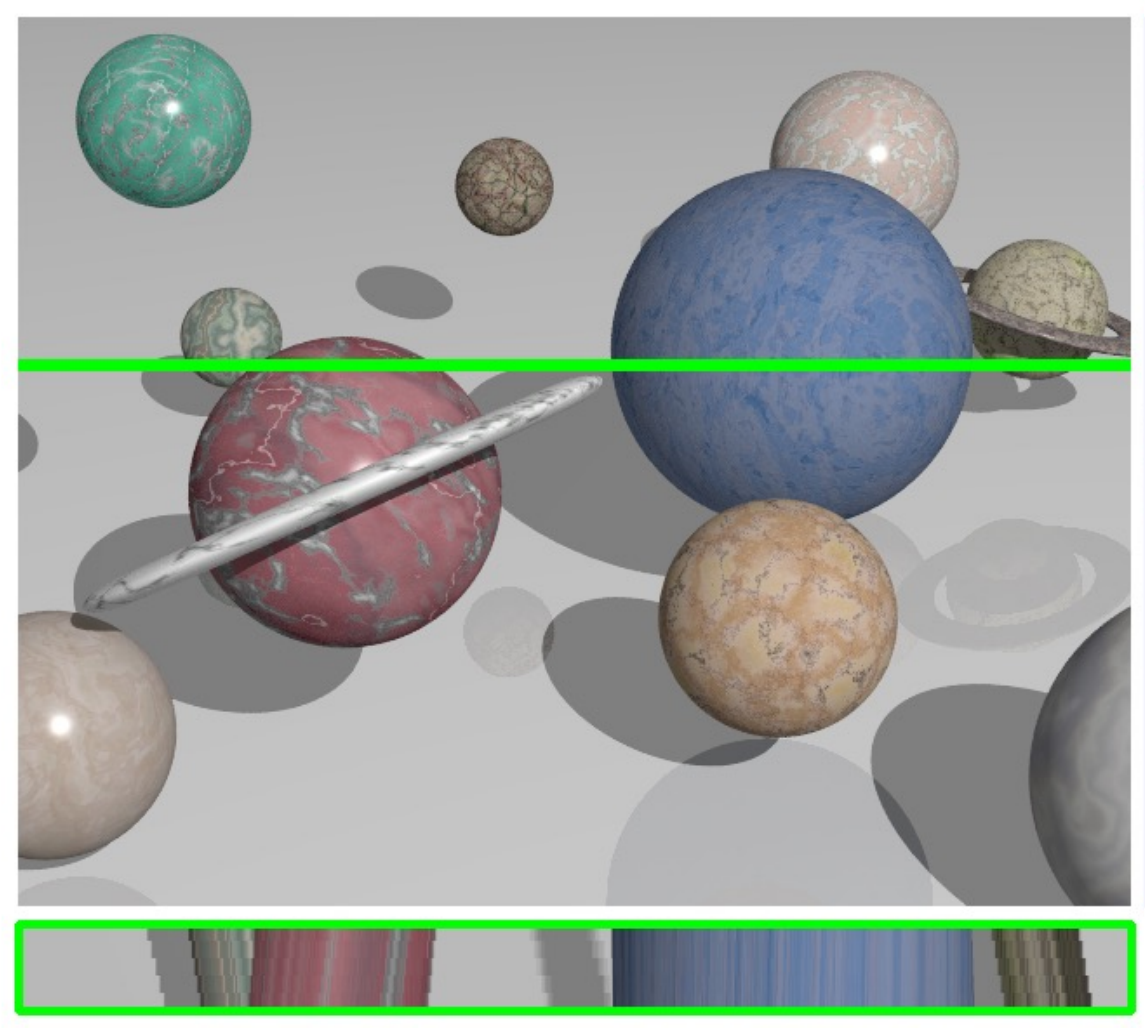}
\centerline{Ground truth}\\
\centerline{(50-th frame)}
\vspace{16.5mm}
\end{minipage}
\begin{minipage}[b]{0.16\hsize}
\centering
\includegraphics[width=.95\linewidth]{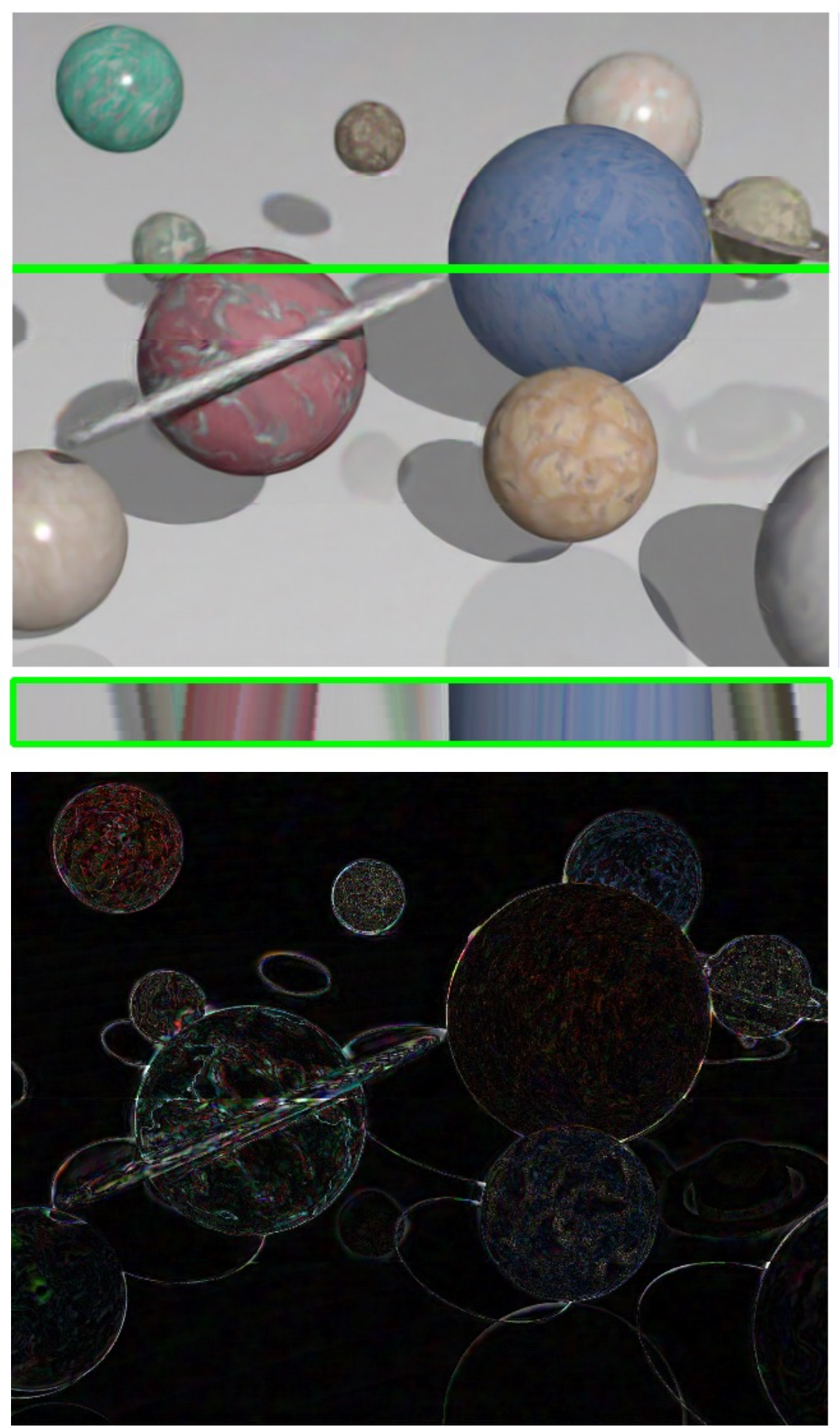}
\centerline{\textbf{A+P~(ours)}}
\end{minipage}
\begin{minipage}[b]{0.16\hsize}
\centering
\includegraphics[width=.95\linewidth]{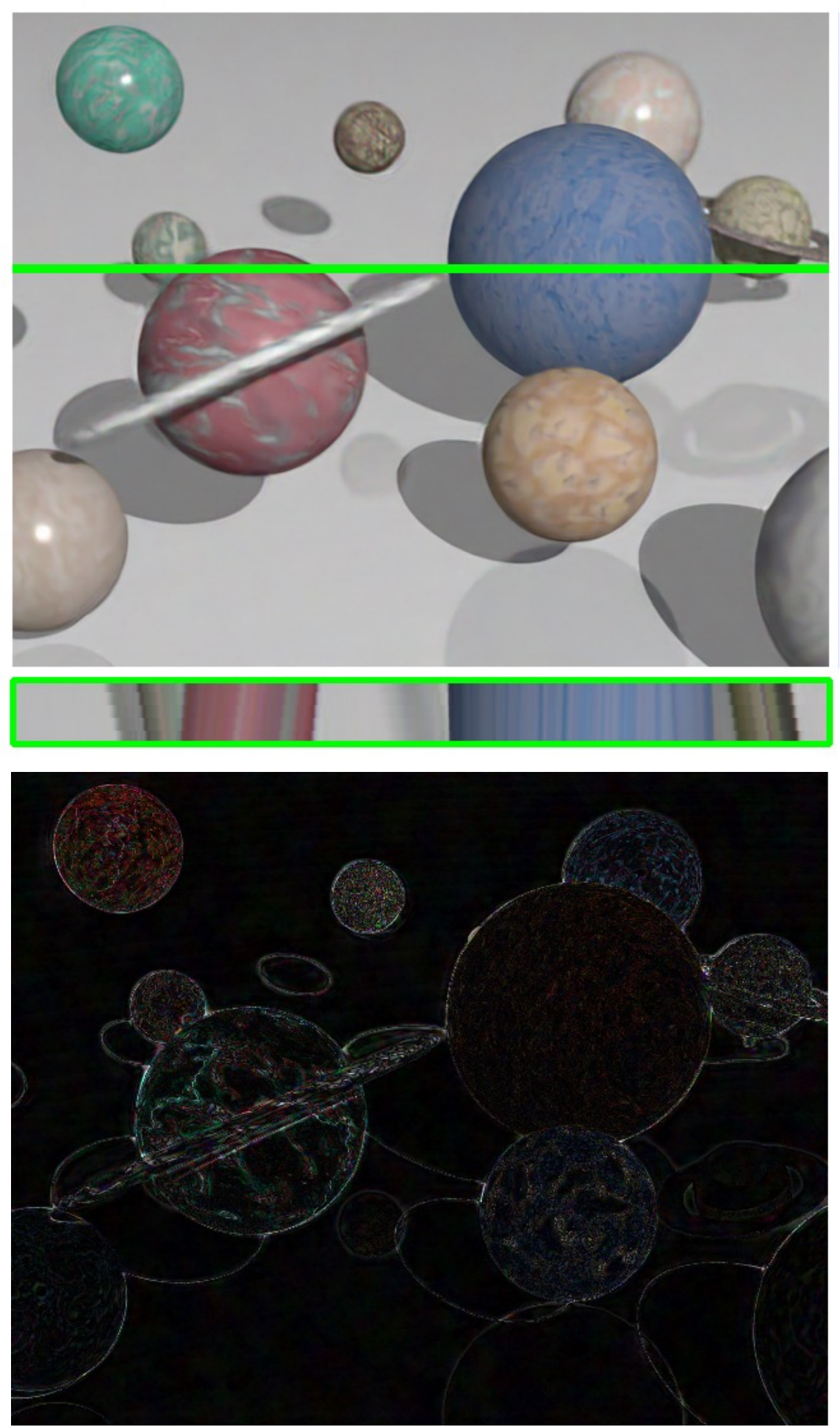}
\centerline{Free5D}
\end{minipage}
\begin{minipage}[b]{0.16\hsize}
\centering
\includegraphics[width=.95\linewidth]{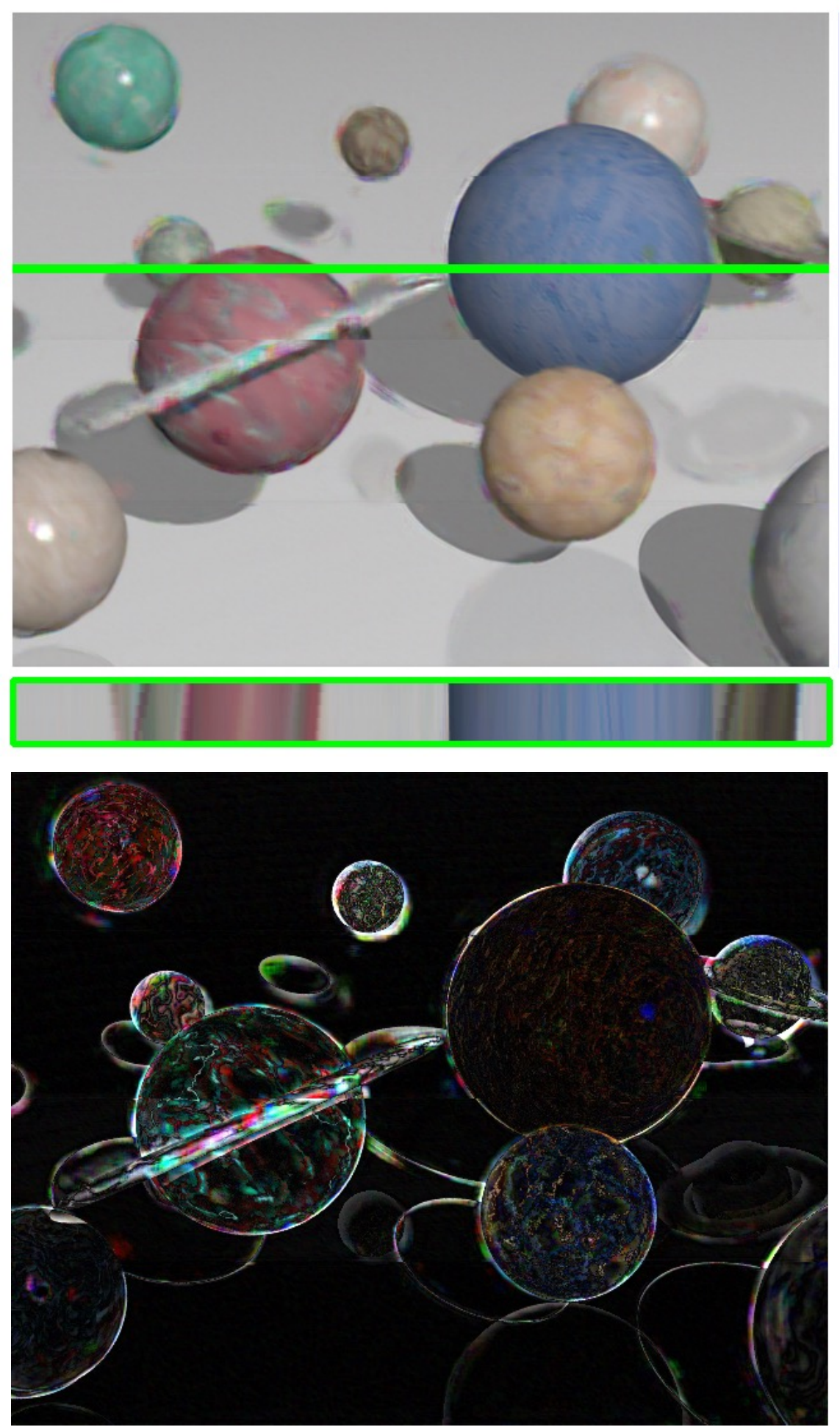}
\centerline{A-only}
\end{minipage}
\begin{minipage}[b]{0.16\hsize}
\centering
\includegraphics[width=.95\linewidth]{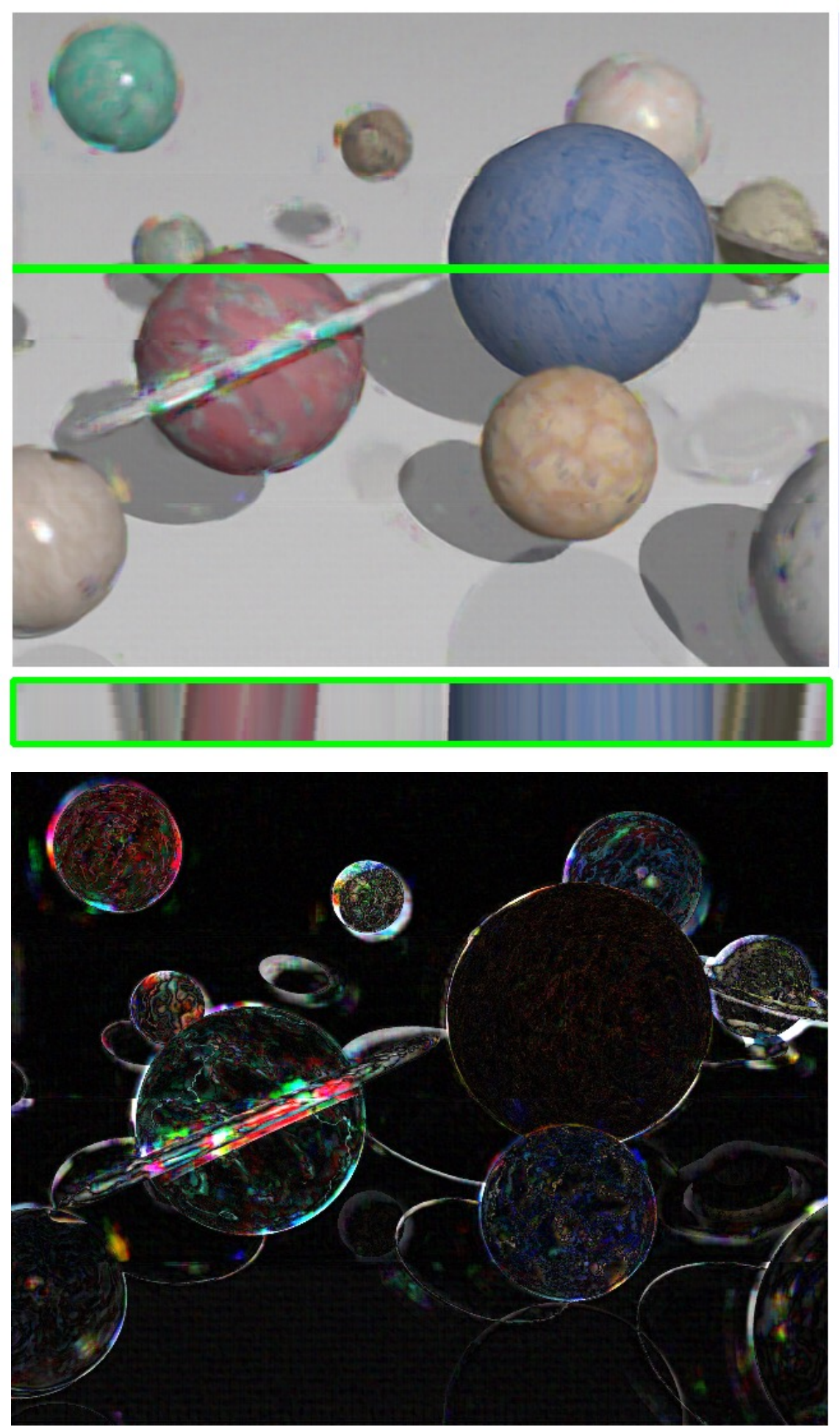}
\centerline{P-only}
\end{minipage}
\begin{minipage}[b]{0.16\hsize}
\centering
\includegraphics[width=.95\linewidth]{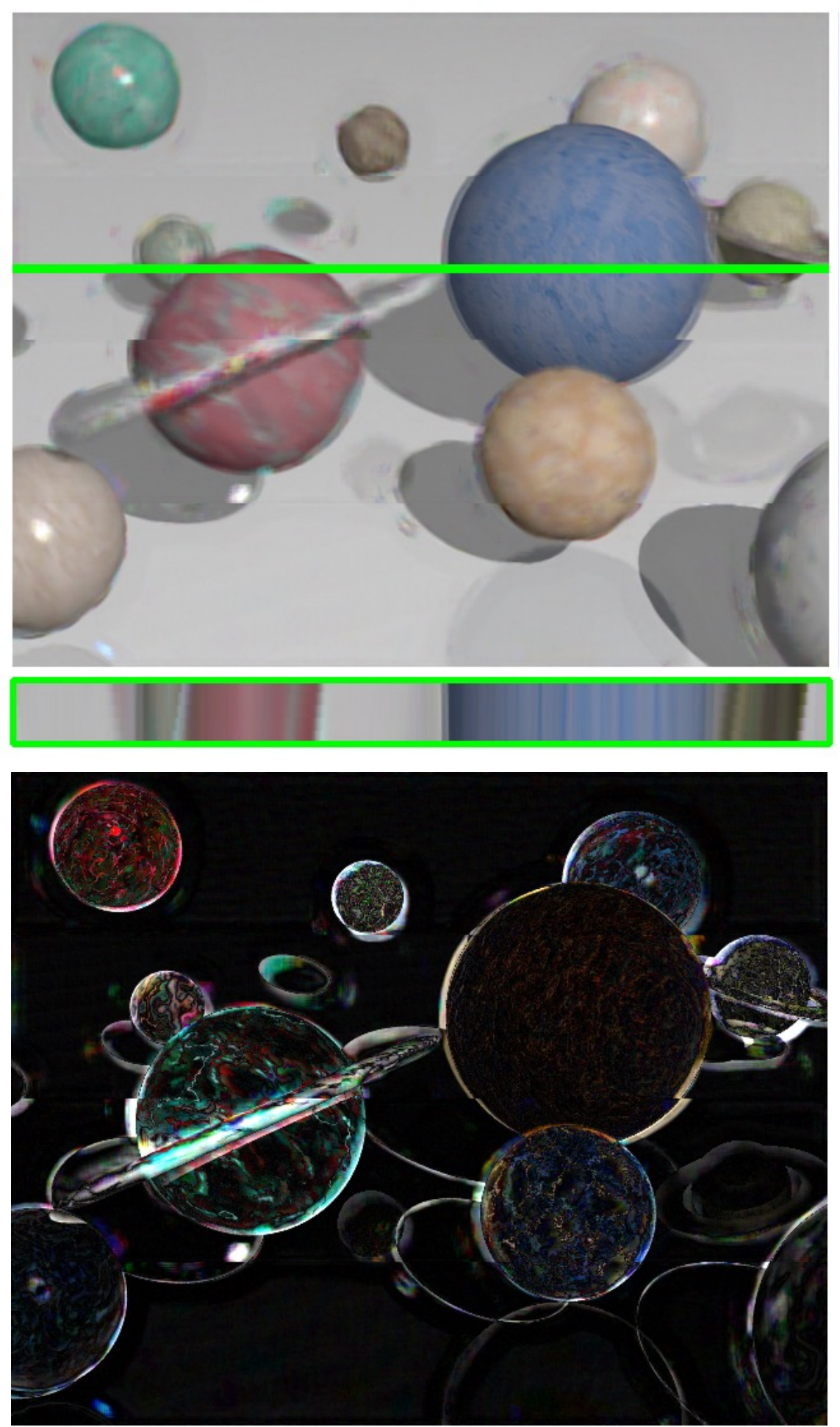}
\centerline{Ordinary}
\end{minipage}
\begin{minipage}[b]{0.16\hsize}
\centering
\includegraphics[width=.95\linewidth]{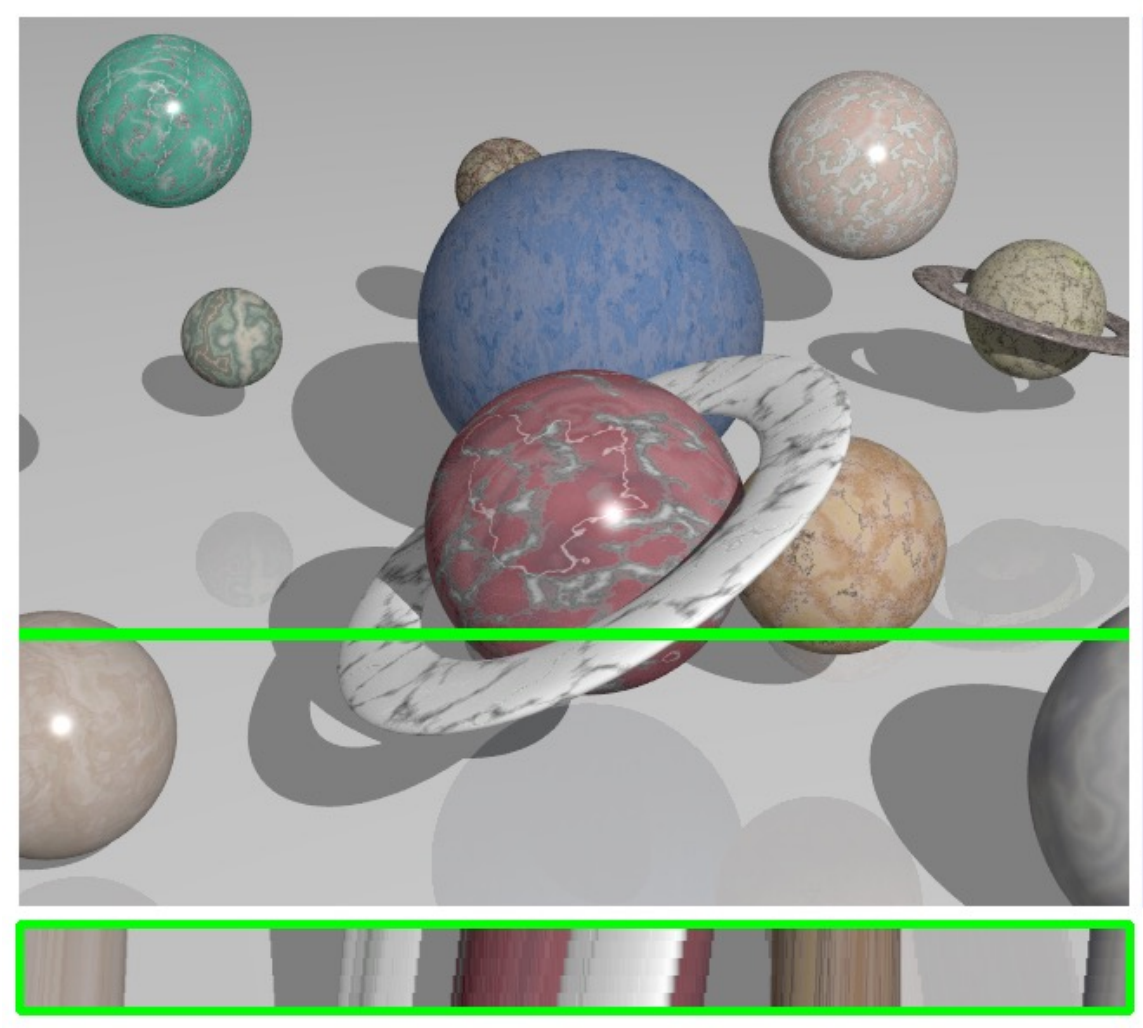}
\centerline{Ground truth}\\
\centerline{(100-th frame)}
\vspace{16.5mm}
\end{minipage}
\begin{minipage}[b]{0.16\hsize}
\centering
\includegraphics[width=.95\linewidth]{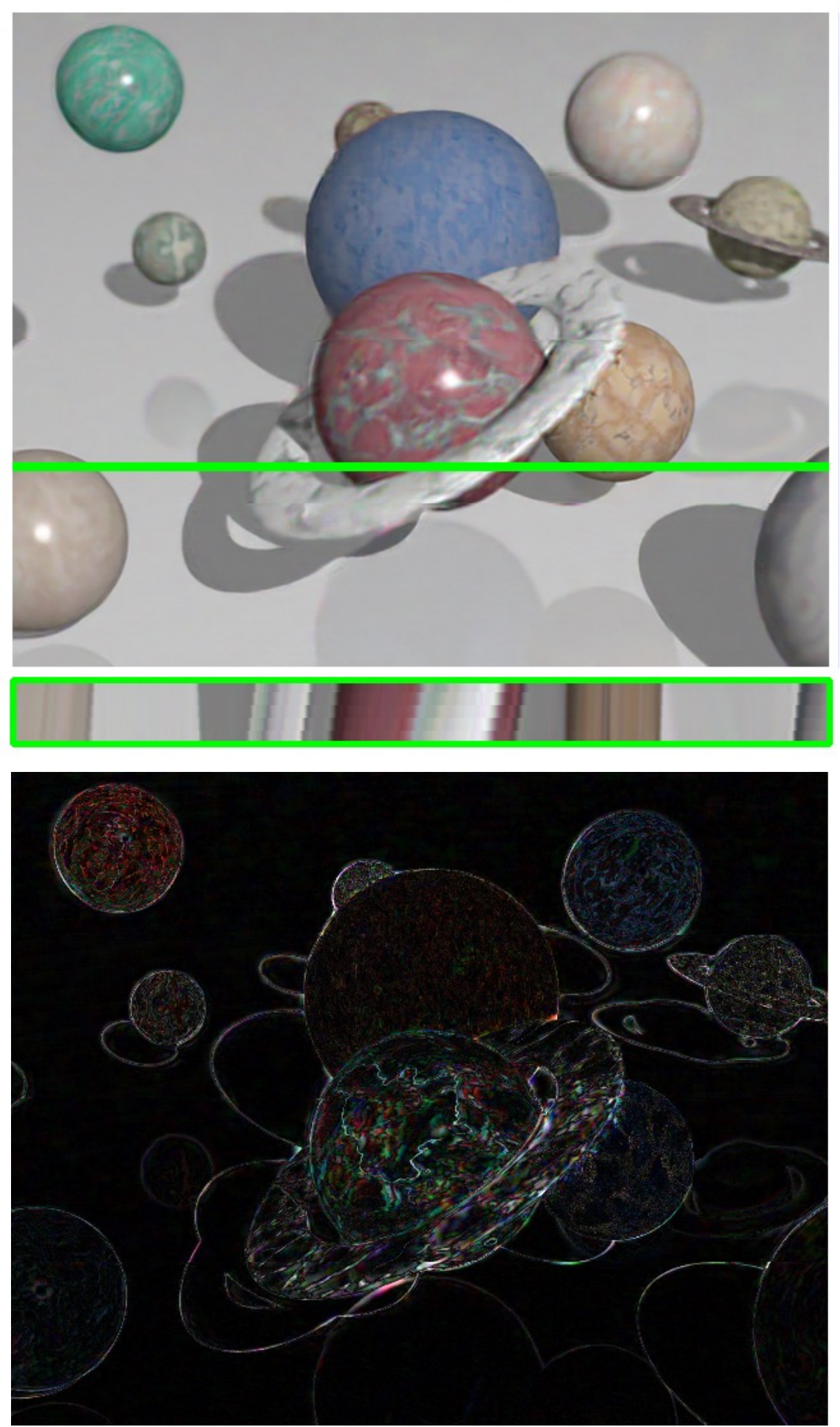}
\centerline{\textbf{A+P~(ours)}}
\end{minipage}
\begin{minipage}[b]{0.16\hsize}
\centering
\includegraphics[width=.95\linewidth]{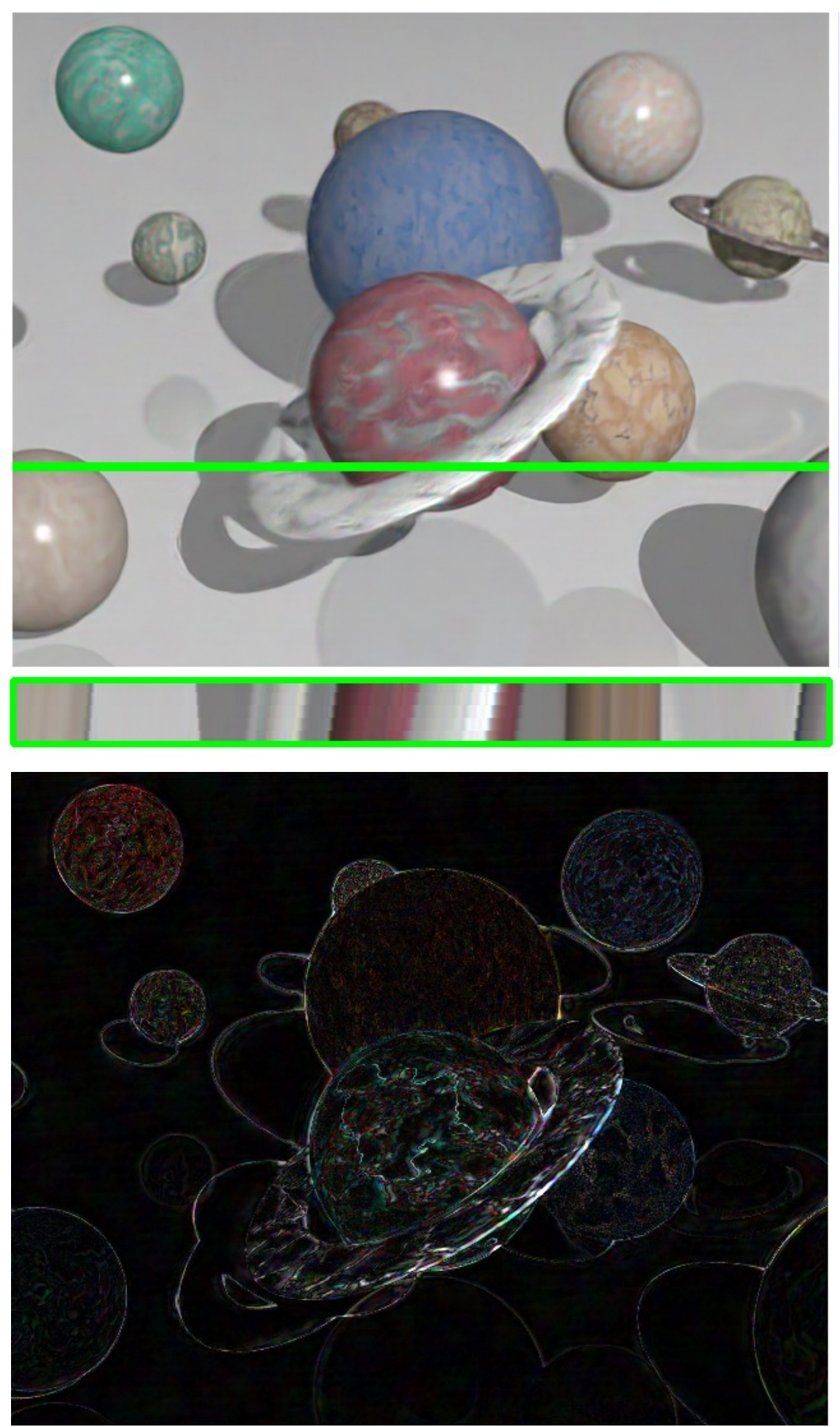}
\centerline{Free5D}
\end{minipage}
\begin{minipage}[b]{0.16\hsize}
\centering
\includegraphics[width=.95\linewidth]{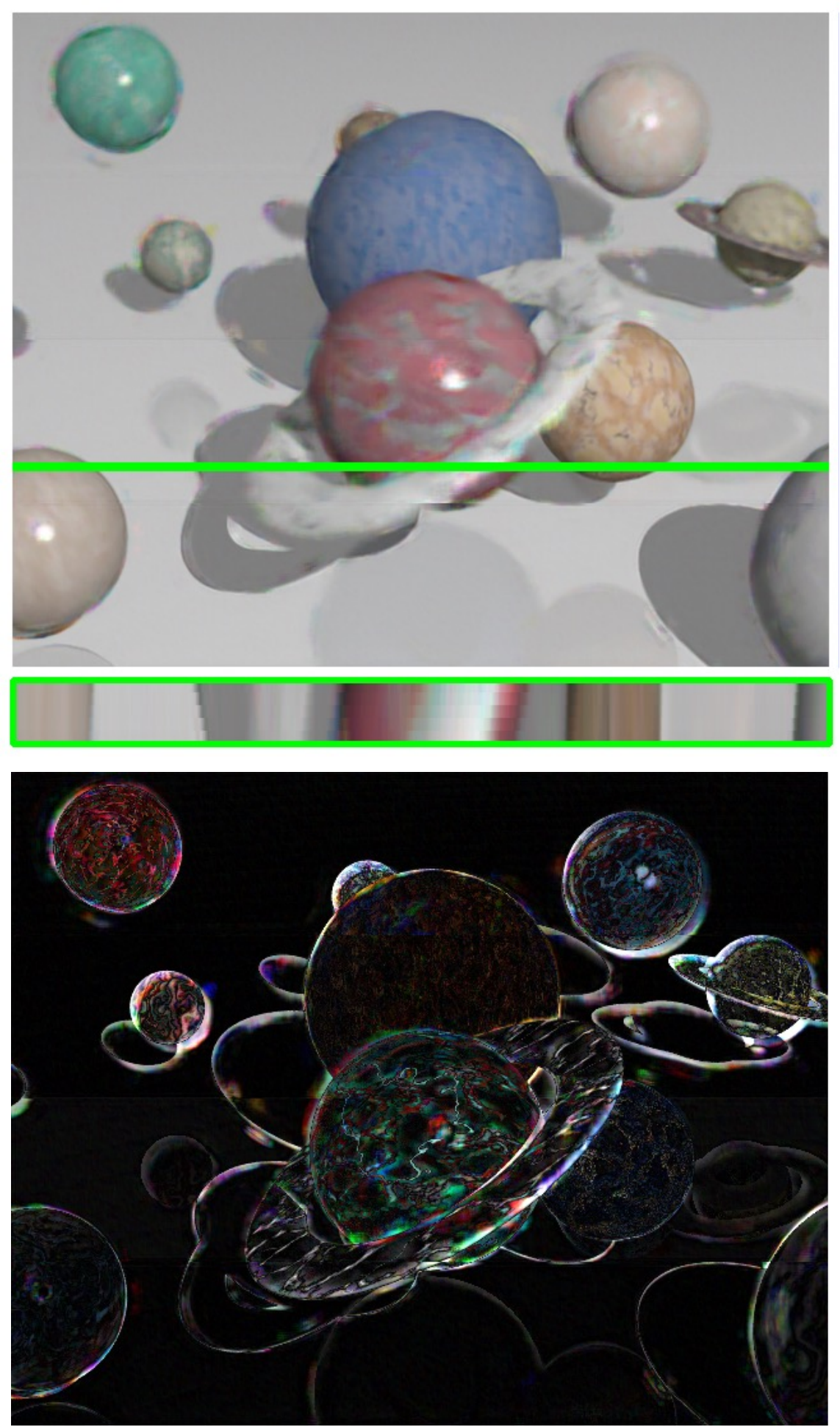}
\centerline{A-only}
\end{minipage}
\begin{minipage}[b]{0.16\hsize}
\centering
\includegraphics[width=.95\linewidth]{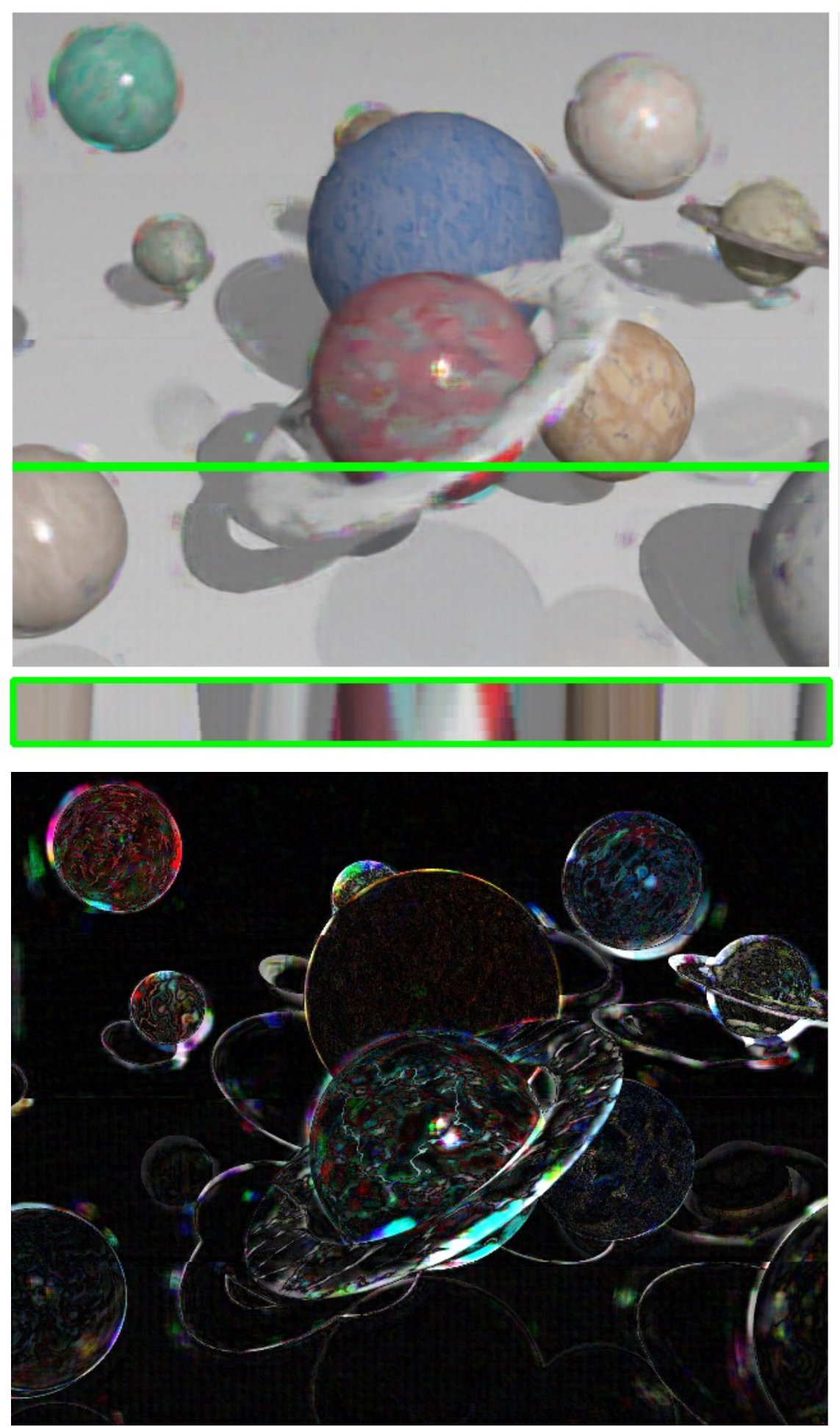}
\centerline{P-only}
\end{minipage}
\begin{minipage}[b]{0.16\hsize}
\centering
\includegraphics[width=.95\linewidth]{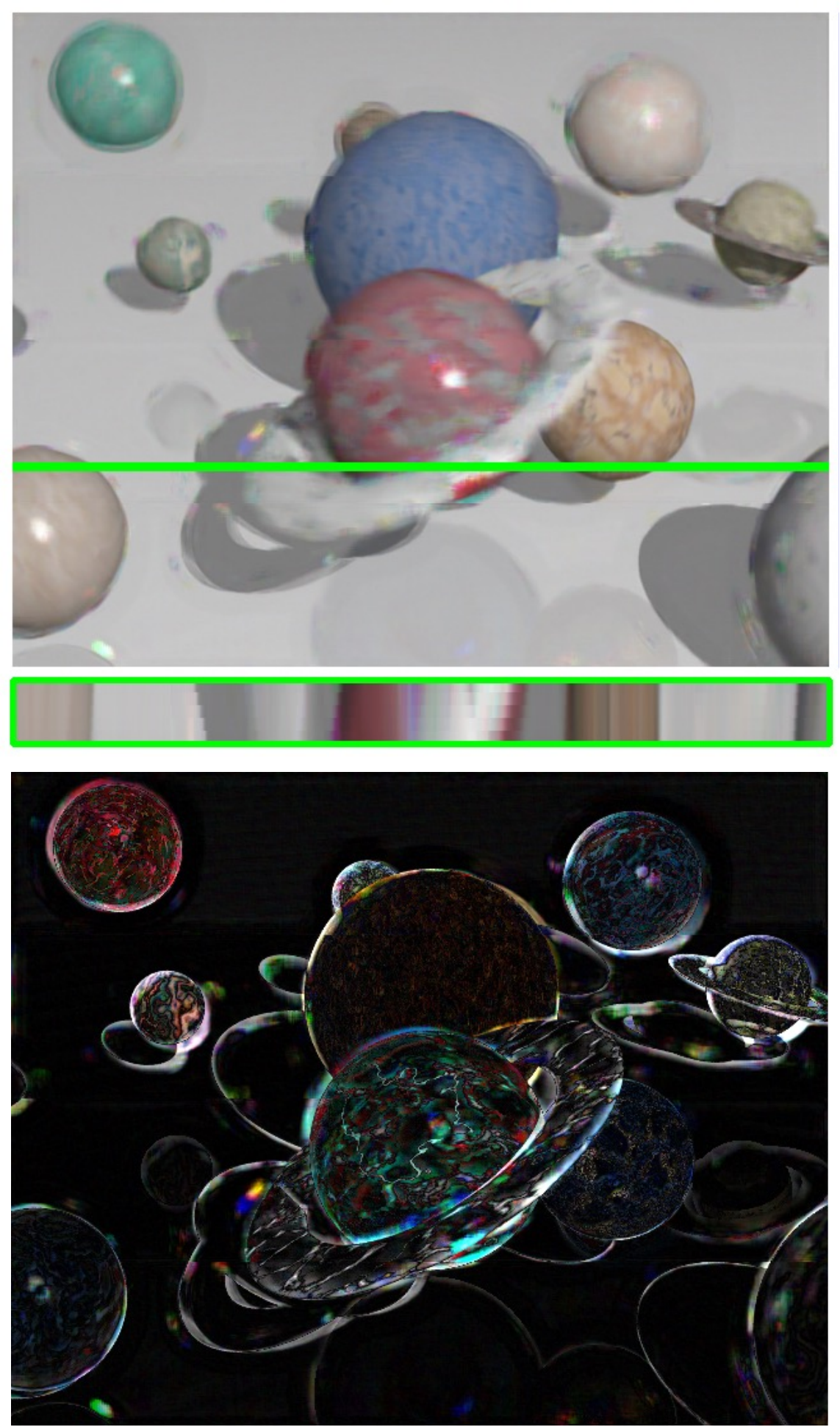}
\centerline{Ordinary}
\end{minipage}
\caption{Visual results of our method (A+P), Free5D (ideal case), and three ablation cases (A-only, P-only, and Ordinary). Reconstructed top-left views are accompanied with horizontal EPIs along green lines and differences from ground truth ($\times 3$ brightness).}
\label{fig:diff}
\end{figure*}

\begin{figure}[t]
\centering
\includegraphics[width=.85\linewidth]{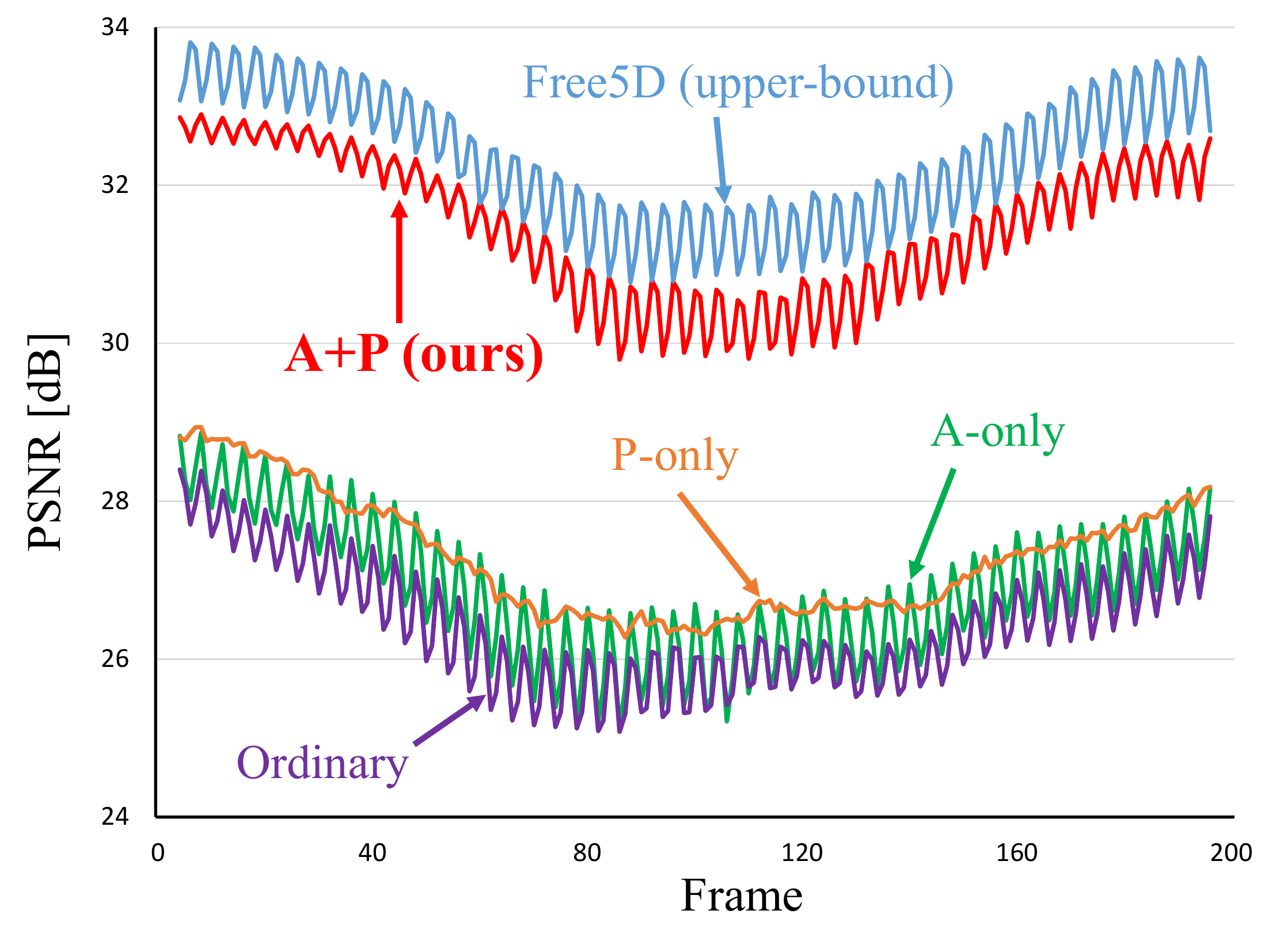}
\caption{Quantitative reconstruction quality over time for our method (A+P), Free5D (ideal case), and three ablation cases (A-only, P-only, and Ordinary).}
\label{fig:graf}
\end{figure}

For quantitative evaluation, we used a computer generated light field with $5 \times 5$ viewpoints over 200 temporal frames, which was rendered from \textit{Planets} scene provided by Sakai et al.~\cite{Sakai_2020_ECCV}.~\footnote{https://www.fujii.nuee.nagoya-u.ac.jp/Research/CompCam/} Figure~\ref{fig:diff} visualizes several reconstructed views (at the top-left viewpoint), horizontal epipolar plane images (EPIs) along the green lines, and the differences from the ground truth ($\times 3$ pixel values). The average peak signal-to-noise ratio (PSNR) values over the 25 viewpoints are plotted along the temporal frames in Fig.~\ref{fig:graf}. 

As observed from these results, our method clearly outperformed the other variants and even achieved quality close to the ideal Free5D case. Meanwhile, A-only and P-only resulted in poor reconstruction quality, showing their insufficiency as coding methods. Moreover, the poor result from Ordinary case indicated that although implicit scene priors were learned from the training dataset, they alone were insufficient for high-quality reconstruction. In contrast, the success of our method can be attributed to the elaborated coding method that was simultaneously applied on the aperture and imaging planes, which helped effectively embed the original 5-D information into a single observed image. However, the reconstruction quality of our method exhibited small fluctuations over time. This was closely related to the fact that four time units (temporal frames) were processed as a group. Moreover, our method did not include mechanisms that could explicitly encourage the temporal consistency, which will be addressed in the future work. 

\begin{figure*}[t]
\centering
\begin{minipage}[b]{0.42\hsize}
\centering
\includegraphics[width=.85\linewidth]{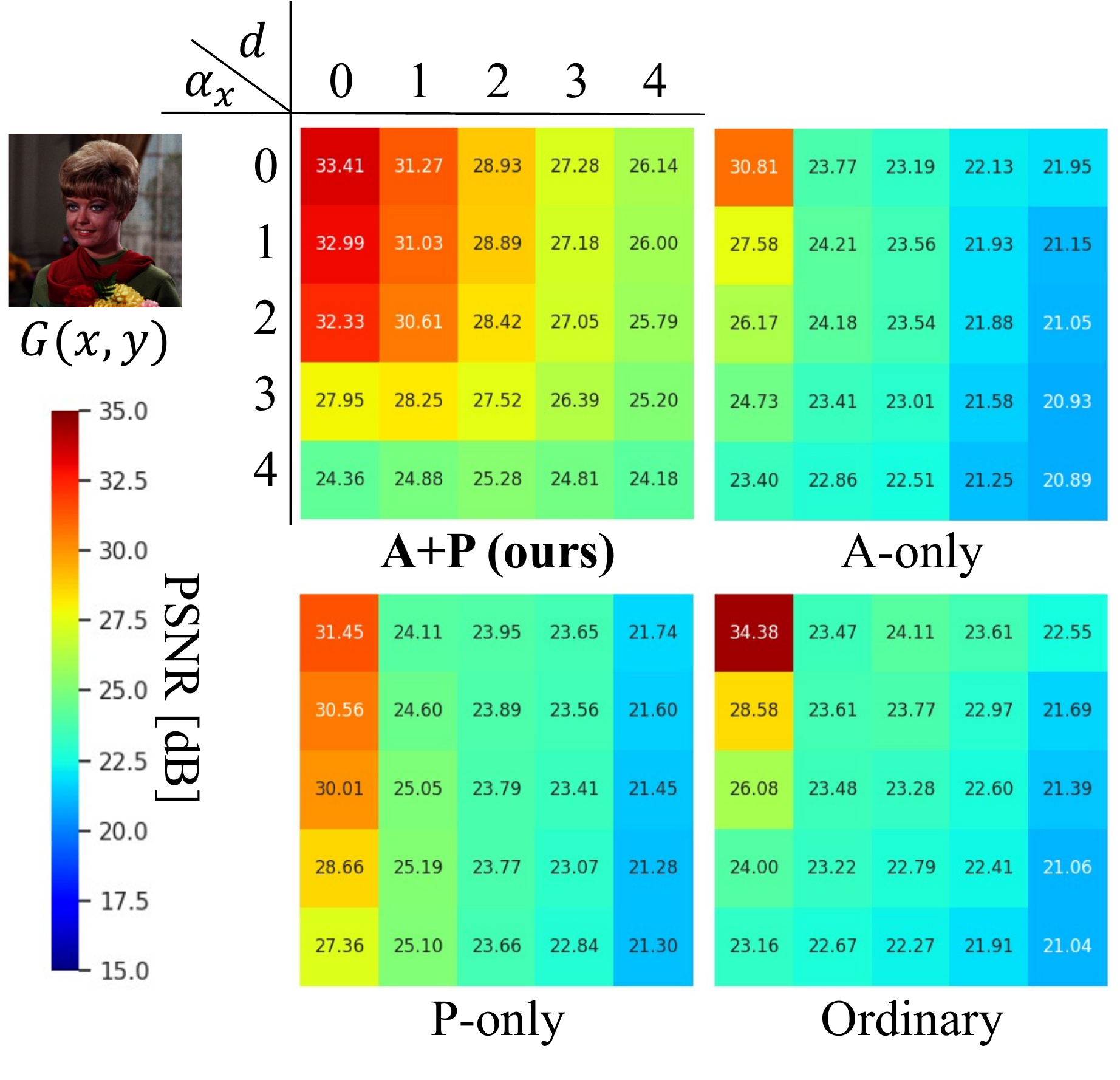}
\caption{Performance evaluation against various motion and disparity values on primitive plane scene.}
\label{fig:working_range}
\end{minipage}
\begin{minipage}[b]{0.07\hsize}
\vspace{30mm}
\end{minipage}
\begin{minipage}[b]{0.50\hsize}
\centering
\includegraphics[width=.85\linewidth]{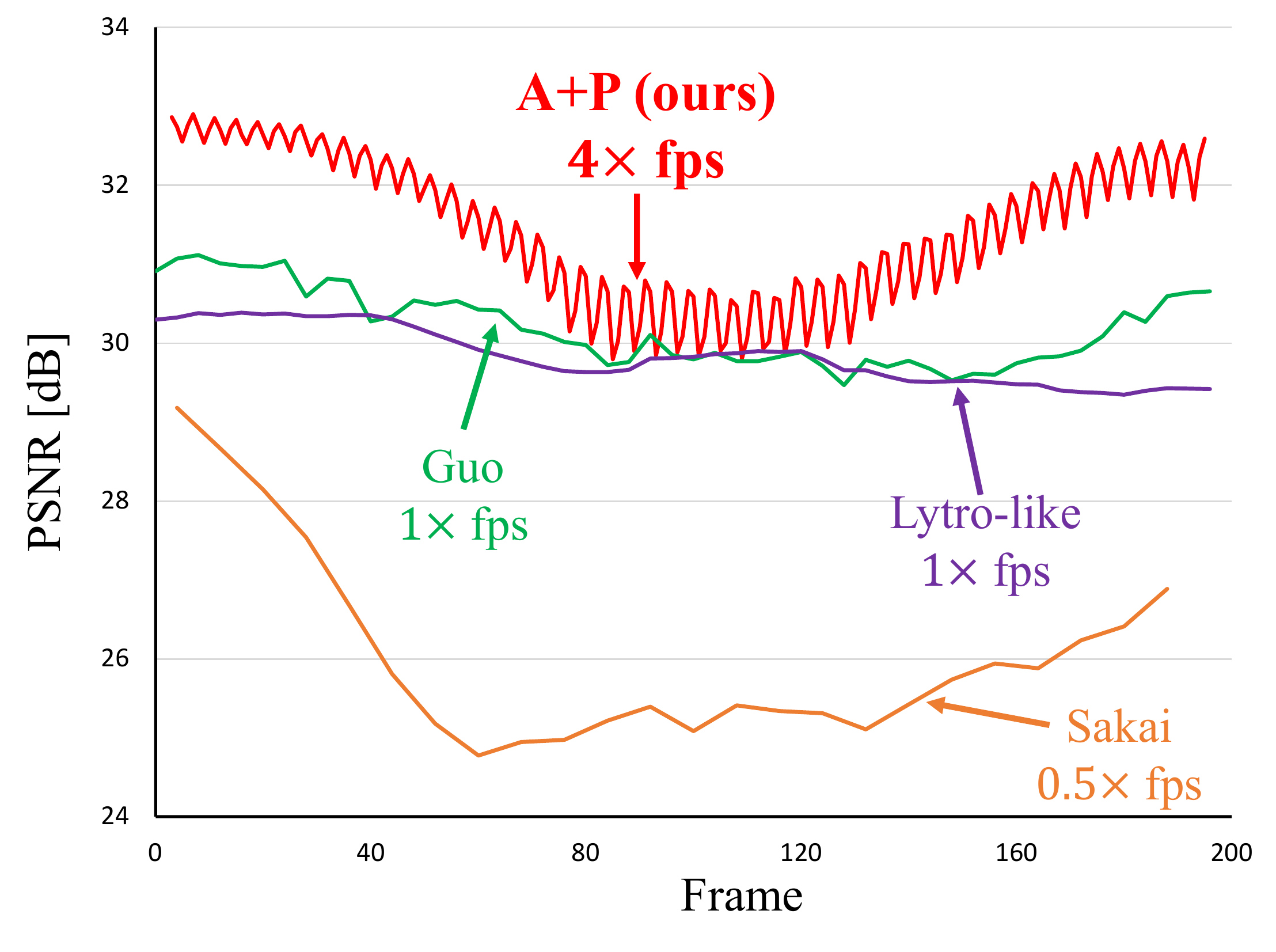}
\caption{Quantitative quality over time compared against other methods (Guo et al.~\cite{Guo_2020_ECCV}, Sakai et al.~\cite{Sakai_2020_ECCV}, and Lytro-like camera).}
\label{fig:guo_vshape}
\end{minipage}
\end{figure*}

\textbf{Working range analysis}. We also evaluated the effective working range against motion and disparity using a primitive plane scene. Following Eq.~(\ref{eq:mklf}), we synthesized a dynamic light field over four time units by using a natural image in Fig.~\ref{fig:working_range} (left) as the texture. The average PSNR values obtained with our method (A+P) and the three variants (A-only, P-only, and Ordinary) are shown in Fig.~\ref{fig:working_range}~(right). Obviously, our method (A+P) can cover a wider range of motion/disparity values than the others; P-only performed poorly for $d \neq 0$; A-only and Ordinary did not work well except for $d = \alpha_x = 0$. 

In our method (A+P), the reconstruction quality degraded gradually as the velocity and disparity values increased. This means that large motions/disparities are challenging for our method. The working range for the disparity was mainly determined by the 3-D scene structures contained in the original light-field dataset, while the working range for the velocity was related to the virtual motions we assumed when synthesizing the dynamic dataset from static light fields. Note that our imaging system has densely-located viewpoints (bounded by the aperture) and a high temporal resolution ($4\times$ the frame-rate of the camera); therefore, both the motion and disparity are usually limited within a small range.


\textbf{Comparison with other methods}. We finally compared our method against three other methods. The first two methods \cite{Guo_2020_ECCV,Sakai_2020_ECCV} were based on coded-aperture imaging. From Guo et al.~\cite{Guo_2020_ECCV}, we adopted a model where a light field for each time unit was reconstructed from a single observed image, which resulted in frame-by-frame observation and light-field reconstruction at the same frame rate as the camera. The method of Sakai et al.~\cite{Sakai_2020_ECCV} observed three consecutive images over time, and reconstructed a light field for the central time. The light field was reconstructed for every two frames (at $0.5\times$ the frame rate) of the camera. We retrained Guo et al.'s and Sakai et al.'s on the same dataset as ours until convergence. In addition, we simulated a Lytro-like camera, where each of the $5\times 5$ views was captured with the $1/5 \times 1/5$ spatial resolution at the same frame rate as the camera. The acquired 5 $\times$ 5 views were upsampled to the original resolution using bicubic interpolation and compared against the ground truth. 

For quantitative evaluation, we used \textit{Planets} assuming the camera's frame rate to be the same as ours; accordingly, in these three methods, image acquisition was conducted only at every four temporal frames. Note that only our method can obtain the light field at $4\times$ the frame rate of the camera, and thus, this comparison only serves as a reference. The average PSNR values over time are shown in Fig.~\ref{fig:guo_vshape}. The method of Sakai et al.~\cite{Sakai_2020_ECCV} failed to follow the fast scene motions, resulting in poor reconstruction quality. The method of Guo et al.~\cite{Guo_2020_ECCV} reconstructed a finely textured but geometrically inconsistent result, whereas the Lytro-like camera produced a geometrically consistent but blurred result. Our method achieved the best reconstruction quality with $\times 4$ finer temporal resolution than the camera. 

Please refer to the supplementary material for more detailed analysis with different training conditions.

\begin{figure*}[t]
\centering
\begin{minipage}[b]{0.35\hsize}
\centering
\includegraphics[width=.95\linewidth]{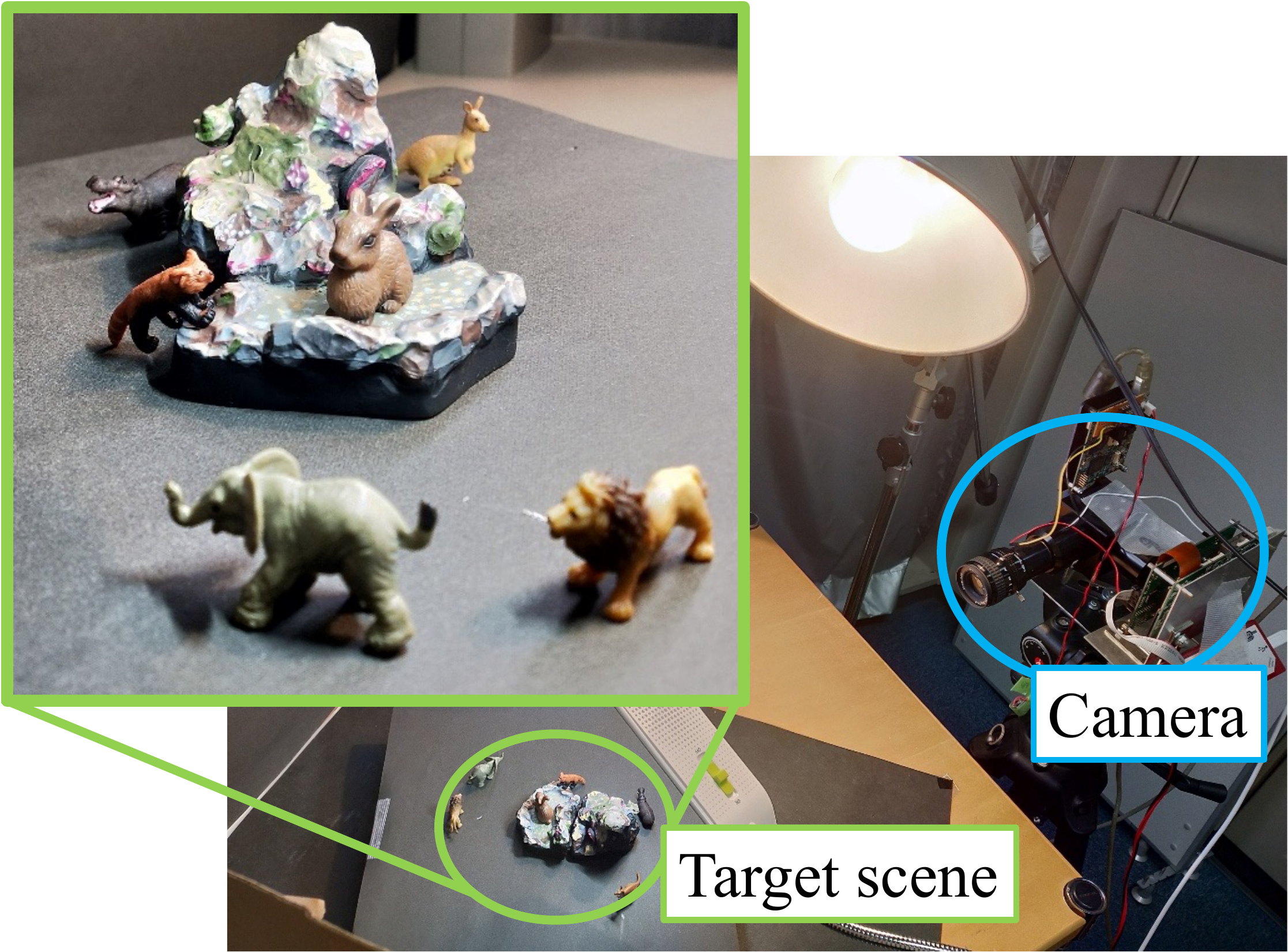}
\centerline{Experimental setup}
\end{minipage}
\begin{minipage}[b]{0.64\hsize}
\centering
\includegraphics[width=.95\linewidth]{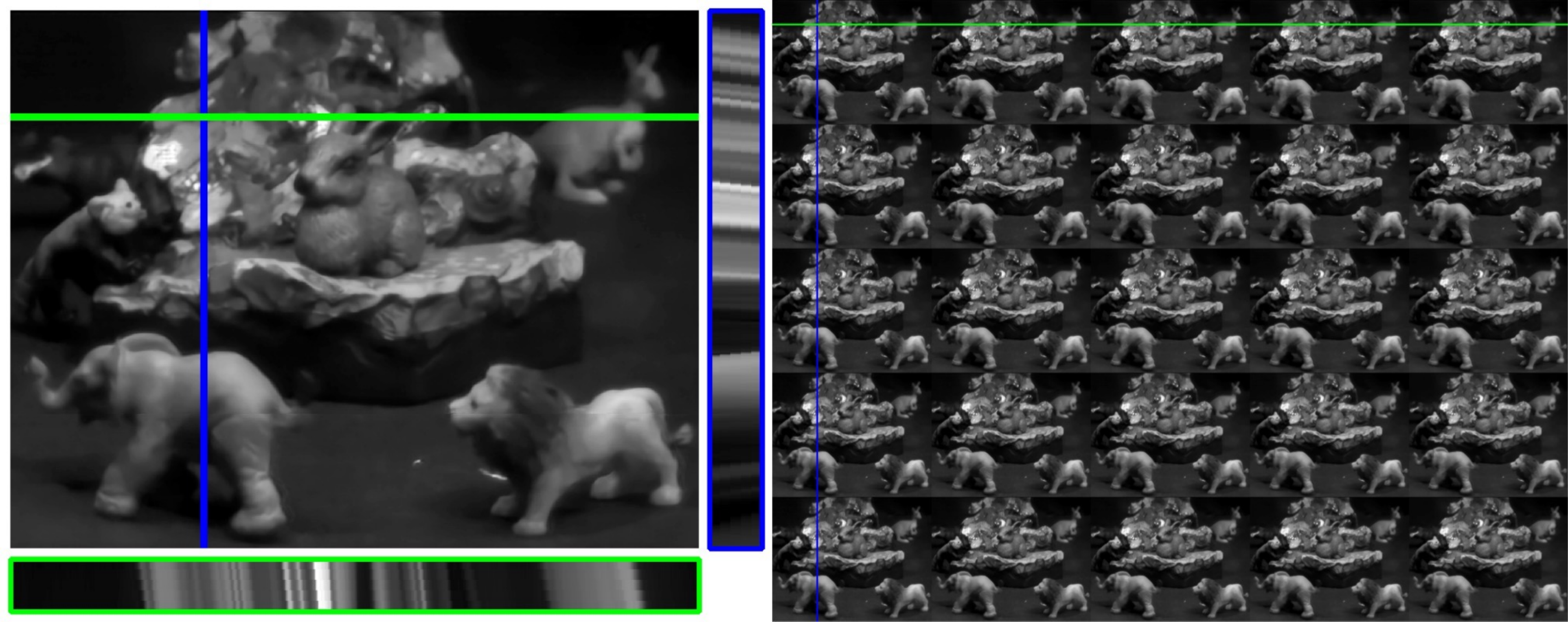}
\centerline{Reconstructed light field}
\end{minipage}
\caption{Experiment using our prototype camera: experimental setup (left) and reconstructed top-left view accompanied by two EPIs along green and blue lines (center), and reconstructed light field with $5\times 5$ views (right).}
\label{fig:real_cam}
\end{figure*}

\subsection{Experiment Using Camera Prototype}
\label{subsec:physical}
We prepared a target scene by using several objects (miniature animals) placed on an electronic turntable, which produced motions in various directions. Our prototype camera was used to capture the scene at 14.7 fps, from which we reconstructed the dynamic light field at 58.8 fps (4 temporal frames from each exposed image). The reconstructed light field had 5 $\times$ 5 views, each with the full-sensor resolution (656 $\times$ 512 pixels) for each time unit. Our experimental setup and a part of the results are shown in Fig.~\ref{fig:real_cam}. The reconstructed light field exhibited natural motions over time and consistent parallaxes among the viewpoints (refer to the supplementary video).

\section{Conclusions}
\label{sec:conclusions}
We proposed a method for compressively acquiring a dynamic light field (a 5-D volume) through a single-shot coded image (a 2-D measurement). Our method was embodied as a camera that synchronously applied aperture coding and pixel-wise exposure coding within a single exposure time, combined with a deep-learning-based algorithm for light-field reconstruction. The coding patterns were jointly optimized with the reconstruction algorithm, so as to embed as much of the original information as possible in a single observed image. Experimental results showed that by using a single camera alone, our method can successfully acquire a dynamic light field with $5 \times 5$ views at $4\times$ the frame rate of the camera. We believe this is a significant advance in the context of compressive light-field acquisition, which will motivate the computational photography community to investigate further. Our future work will include improvement on the network design for better reconstruction quality and generalization to different configurations concerning the number of views and the number of time units included in a single exposure time.

\noindent
\textbf{Acknowledgement:} Special thanks go to Yukinobu Sugiyama and Kenta Endo at Hamamatsu Photonics K.K. for providing the image sensor.

{\small
\bibliographystyle{ieee_fullname}
\bibliography{refs}
}

\end{document}